\documentclass[aps,prb,twocolumn,showpacs,amsmath,amssymb]{revtex4}
\usepackage{graphicx}
\usepackage{bm}
\def\be{\begin{equation}}
\def\ee{\end{equation}}
\def\bea{\begin{eqnarray}}
\def\eea{\end{eqnarray}}

\begin{document}

\title{Fluctuation Statistics in Networks: a Stochastic Path Integral Approach}
\author{Andrew N. Jordan}
\email{Andrew.Jordan@physics.unige.ch}
\author{Eugene V. Sukhorukov}
\author{Sebastian Pilgram}
\affiliation{D\'epartement de Physique Th\'eorique, Universit\'e de Gen\`eve,
CH-1211 Gen\`eve 4, Switzerland}

\begin{abstract}
We investigate the statistics of fluctuations in a classical stochastic network
of nodes joined by connectors. The nodes carry generalized charge that may be
randomly transferred from one node to another. Our goal is to find the time
evolution of the probability distribution of charges in the network.  The
building blocks of our theoretical approach are (1) known probability
distributions for the connector currents, (2) physical constraints such as
local charge conservation, and (3) a time-scale separation between the slow
charge dynamics of the nodes and the fast current fluctuations of the
connectors.  We integrate out fast current fluctuations and derive a
stochastic path integral representation of the evolution operator for the slow
charges. The statistics of charge fluctuations may be found from the
saddle-point approximation of the action.  Once the probability distributions
on the discrete network have been studied, the continuum limit is taken to
obtain a statistical field theory. We find a correspondence between 
the diffusive field theory and a Langevin equation with 
Gaussian noise sources, leading
nevertheless to non-trivial fluctuation statistics.  To complete our theory,
we demonstrate that the cascade diagrammatics, recently introduced by Nagaev,
naturally follows from the stochastic path integral. By generalizing the
principle of minimal correlations, we extend the diagrammatics to calculate
current correlation functions for an arbitrary network.  One primary
application of this formalism is that of full counting statistics (FCS), the
motivation for why it was developed in the first place. We stress however,
that the formalism is suitable for general classical stochastic problems as an
alternative approach to the traditional master equation or Doi-Peliti
technique.  The formalism is illustrated with several examples: both
instantaneous and time averaged charge fluctuation statistics in a mesoscopic
chaotic cavity, as well as the FCS and new results for a generalized diffusive
wire.
\end{abstract}

\maketitle

\section{Introduction}
\label{intro}

Consider an exclusive night-club with a long line at the entrance.  A bouncer
is at the front of the line to keep out the rif-raf.  At every time step, a
person is accepted inside the club with probability $p$, or rejected with
probability $1-p$. Inside the club, people stay for a while and eventually
leave.  At every time step, the probability a person leaves is $q$.  We want
to answer a question such as ``what is the probability that $Q$ people leave
the club after $t$ time steps?''.

Assuming that $p$ and $q$ remain constant, the situation is simple and we can
easily solve the relevant probabilistic problem.  However, in realistic
situations this rarely happens: the management wants to make money.  If the
club is almost empty, they instruct the bouncer to be less discriminating,
while if the club is almost full, the bouncer is to be more discriminating.
Thus, $p$ becomes a function of the number of people in the club.  People will
be more likely to leave if the club is very crowded, so $q$ is also a function
of the number of people inside the club.  The problem posed now is much more
difficult because of the presence of feed-back: the elementary processes
change in response to the cumulative effect of what they have accomplished in
the past.

This simple example captures all the basic features of the problems we wish to
consider.  Although the example was given with people, the actors in the
probability game may be any quantity such as charge, energy, heat or
particles, which we will refer to simply as generalized charge.  Similarly,
the night club can be a mesoscopic chaotic cavity,\cite{BB} a birth-death
process,\cite{b-d} a biological membrane channel,\cite{examples} etc.

Historically, general stochastic problems are solved with the master equation.
The time rate of change of the probability to be in a particular state is
given in terms of transition rates to other states.  This approach has had
great success and leads naturally to the Fokker-Planck and Langevin
equations.\cite{gardiner} However, once the master equation is given, the
solution is often quite difficult to obtain.

This paper takes a different approach.  Rather than beginning with a master
equation describing the probability of all processes happening in a unit of
time, we make several assumptions from which we can reformulate the problem.
Although these assumptions limit the applicability of the theory, when they
apply, the problems are much easier to solve. The assumptions are:
\begin{itemize}
\item The system we are interested in is a composite system made out of
constituent parts. In the night club example, the system is made up of three
physical regions: outside the front door, the interior of the club, and
outside the back door.  The decomposition of a larger system into smaller
interacting parts is only meaningful for us if there is a separation of time
scales.  This means that the charge inside the constituent parts changes on a
slower time scale than the fluctuations at the boundaries. In the night club
example, this simply means that the average time a person spends in the club
will be much longer than the typical time needed to enter the door.
\item Taken alone, the parts of the composite system have a finite number of
simple properties or parameters.  The only property of the night club that was
relevant for the problem was the total number of people in it at any given
time.  The important element of the line out in front is that it never runs
out.  All other details are irrelevant.
\item In the limit where all parts of the network are very large (so that the
elementary transport processes do not affect themselves in the short-run), the
transport probability distributions between elements are known.  In the night
club example, the probability of getting $Q$ people through the front door
after $t$ time steps (given a constant, large number of people inside) is easy
to find, because we have assumed that the elementary probability $p$ does not
change from trial to trial. The transport probability distribution is simply
the binomial distribution \cite{gardiner} where the probability $p$ is a
function of the (approximately unchanging) number of people inside.  The back
door distribution is obtained in the same way.
\item There are conservation laws that govern the probabilistic processes.
No matter what probability distributions we have, there are certain rules that
must be obeyed.  The net number of people that enter, stay, and leave the club
must be a constant.  This means that the time rate of change of the club's
occupancy is given by the people-current in minus the people-current out.  The
people in the line outside are a special case.  There is in principle always a
replacement, so moving one person inside the club doesn't affect the
properties of the line.  
\end{itemize}

Now, the strategy is to use this information as the starting point to find
transport statistics for the combined interacting system.  The main result
derived is a path integral expression for the conditional probability (taking
conservation laws into account) for starting and ending with a given amount of
charge at each location after some time has passed.  From this conditional
probability, specific quantities such as transport statistics through the
system, fluctuation statistics of charge at a particular location and the like
may be found.


One primary application of this formalism is that of full counting statistics
(FCS), \cite{LevitovStatistics,Levitov2} the motivation for why it was
developed in the first place.\cite{us} FCS describes the fluctuations of
currents in electrical conductors. It gives the distribution of the
probability that a certain number of electrons pass a conductor in certain
amount of time. Mean current flow and and shot noise\cite{BB} correspond to
the first and second cumulant of this distribution. The full distribution
(defined by all cumulants) provides a full characterization of the transport
properties of a electrical conductor in the long time limit.  In the past, FCS
was mainly addressed with quantum mechanical tools such as the scattering
theory\cite{LevitovStatistics,wire,Levitov2,Muzykantskii1} of coherent
conductors, the circuit theory based on Keldysh Green
functions,\cite{Nazarov1,Belzig1,Bagrets1,Kindermann1} or the nonlinear
$\sigma$ model.\cite{Gefen1} However, a number of works realized that for
semi-classical systems with a large number of conductance channels, shot noise
may be calculated without accounting for the phase coherence of the
electron.\cite{kirillrate,Beenakker1,Langen,CavityMinCorr} These works treat
the basic sources of noise quantum mechanically, but calculate the spread of
the noise throughout the conductor classically. For specific conductors like
diffusive wires and chaotic cavities, this idea has been extended to the
calculation of third and fourth cumulants via the cascade
principle\cite{Nagaev1,CavityKirill} and to the full generating function of
FCS.\cite{DeJong1,us,derrida,derrida2} In the present work, we consider an
abstract model instead of any particular example and develop the mathematical
foundations of the proposed semi-classical procedure to obtain FCS. We
introduce and investigate networks of elements with known transport statistics
and show how the FCS of the entire network can be constructed systematically.

The formalism we present is related to a different approach in non-equilibrium
statistical physics called the Doi-Peliti technique.\cite{dp} The idea is that
once the basic master equation governing the time evolution of probability
distributions is given, it may be interpreted as a Schr\"odinger equation
which may be cast into a second-quantized language.  This quantum problem is
then converted into a quantum mechanical path integral (often obeying bosonic
or fermionic statistics) from which one may take the continuum limit and use a
field theory renormalization group approach with diagrammatic perturbation
expansion.\cite{cardy} This approach is useful in many situations far from
equilibrium and has several parallels to our approach.  It has been pointed
out that this technique is in some sense the classical limit of the quantum
mechanical Keldysh formalism, \cite{ak} the same tool used in the past to
calculate FCS, so this gives another connection with the subject matter we are
concerned with.

There are several advantages of our approach.  First, we skip the master
equation step.  If the probability distributions of the connector fluctuations
are given, we may immediately construct network distributions.  Second, from a
computational view, our formulation of the problem is much simpler than
starting from first principles for situations where the ingredients we need
are available, and results are much easier to obtain than beginning with the
master equation alone.  Thirdly, our formulation also applies to situations
where temporal transition probabilities may be large.  Finally, the
formalism's physical origin is clear, so the needed mathematical objects are
well motivated.

\begin{figure}[th]
\centerline{\includegraphics[width=7cm]{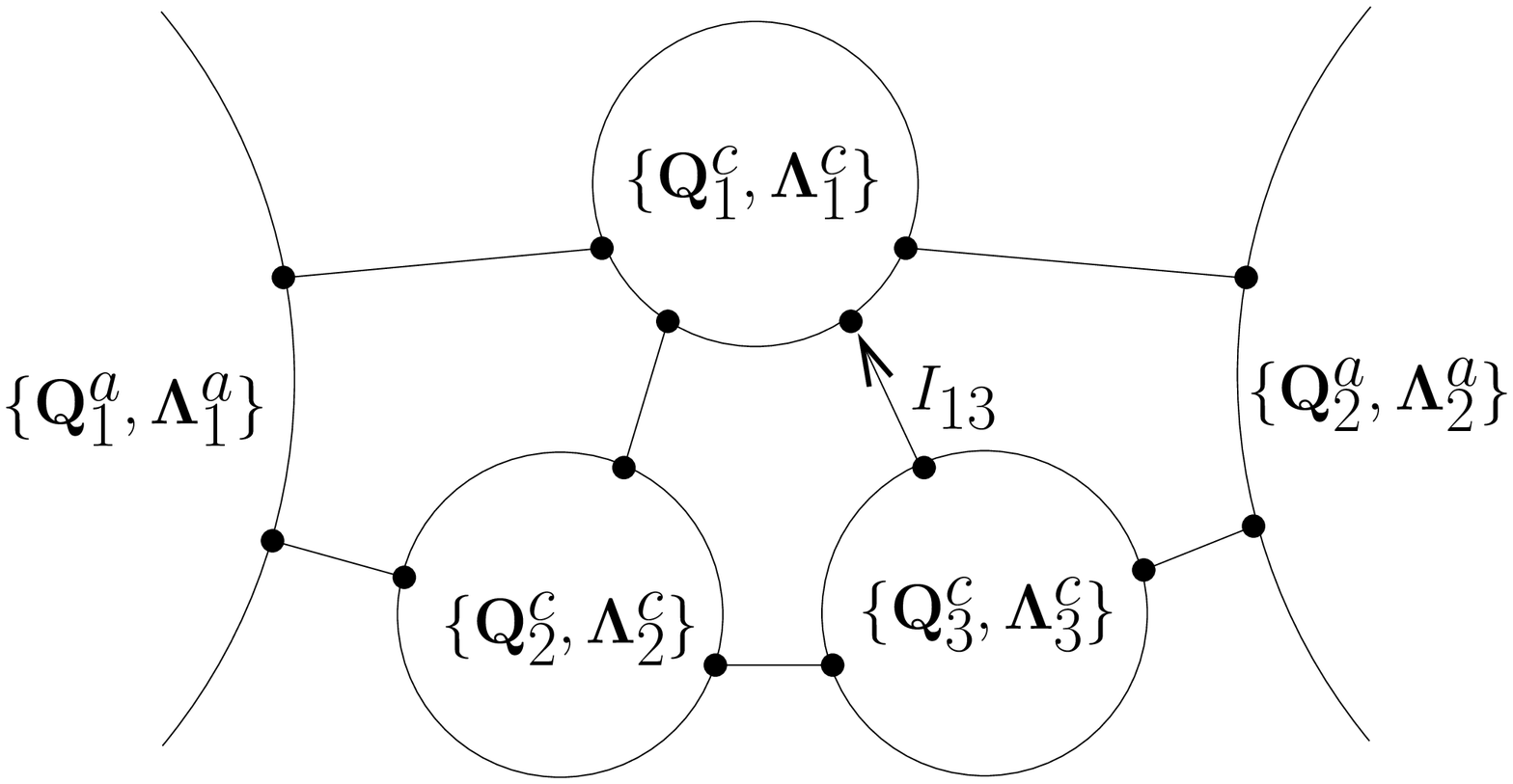}}
\caption{An arbitrary network. Each node has charge and counting variables $\{
{\bf Q}_\alpha, {\bf \Lambda}_\alpha \}$.  The nodes transfer charge via
currents $I_{\alpha\beta}$ through the connectors.  The absorbed counting
fields (${\bf \Lambda}^a_\alpha$) are constants by definition of the absorbed
charges ${\bf Q}^a_\alpha$ (see text).  Each node may have an arbitrary number
of different charge species, ${\bf Q}_\alpha = \{ Q^1_\alpha, Q^2_\alpha,
\ldots, Q^j_\alpha \}$. }
\label{network}
\end{figure} 

The rest of the paper is organized as follows.  In Sec.\ \ref{gfor}, we
introduce and develop the general theory.  After reviewing elements of
probability theory, we derive the stochastic path integral for a network of
nodes as well as explore the relationship to the master equation and
Doi-Peliti formalism.  In Sec.\ \ref{ft}, the continuum limit is taken to
derive a stochastic field theory and link our formalism with the Langevin
equation point of view.  In Sec.\ \ref{cr}, we develop diagrammatics rules to
calculate cumulants of the current distribution as well as current correlation
functions for an arbitrary network.  Sec.\ \ref{app} gives several
applications of the theory to different physical situations.  We solve the
field theory for the mesoscopic wire and demonstrate universality in multiple
dimensions as well as present new results for the conditional occupation
function and probability distribution.  We also consider the problem of charge
fluctuation statistics (both instantaneous and time-averaged) in a mesoscopic
chaotic cavity. Sec.\ \ref{conclusion} contains our conclusions.

\section{General Formalism} 
\label{gfor}

Once we have the basic elements of our theory (the generalized charges), we
must specify some spatial structure that they move around on.  As we noted in
the introduction, the essential structure needed to state the problem are
simply points we refer to as nodes, joined by connectors.  This defines a
network (see Fig.\ \ref{network}).
The state of each node $\alpha$ is described by one (effectively 
continuous) charge $Q_\alpha$,
\cite{multispecies} and ${\bf Q}$ is the charge vector describing the charge
state of the network.  The node's state may be changed by transport: flow of
charges between nodes takes place via the connectors carrying currents
$I_{\alpha\beta}$ from node $\alpha$ to node $\beta$.  The variation of
these charges $Q_{\alpha}$ is given by 
\be 
Q_{\alpha}(t+\Delta t)-Q_{\alpha}(t)=\sum_{\beta}Q_{\alpha\beta}\, ,
\label{cons}
\ee 
where the transmitted charges $Q_{\alpha\beta}(t)=\int_0^{\Delta
t} dt' I_{\alpha\beta}(t+t')$ are distributed according to
$P_{\alpha\beta}(Q_{\alpha\beta}(t))$.  The fact that the
probabilities $P_{\alpha\beta}(t)$ also depend on the charges ${\bf
Q}(t)$ is one source of the difficulty of the problem.

Assuming that the probability distributions $P_{\alpha\beta}$ (which depend
parametrically on the state of nodes $\alpha$ and $\beta$) of the transmitted
charges $Q_{\alpha\beta}$ are known, we seek the time evolved probability
distribution $\Gamma({\bf Q},t)$ of the set of charges ${\bf Q}$ for a given
initial distribution $\Gamma({\bf Q},0)$. In other words, one has to find the
conditional probability (which we refer to as the evolution operator) $U({\bf
Q},{\bf Q}',t)$ such that 
\be \Gamma({\bf Q},t)=\int d{\bf Q}'\, U({\bf
Q},{\bf Q}',t)\Gamma({\bf Q}',0)\, .
\label{gamma}
\ee 
We assume that there is a separation of time scales, $\tau_0 \ll \tau_C$,
between the correlation time of current fluctuations, $\tau_0$, and the slow
relaxation time of charges in the nodes, $\tau_C$.  As we will show in the
next section, this separation of time scales allows us to derive a stochastic 
path integral representation for the evolution operator,
\begin{subequations}
\label{pathint}
\bea 
U({\bf Q}_f,{\bf Q}_i,t) 
&=&\int\! {\cal D}{\bf Q}{\cal D}{\bf \Lambda}
\exp\{S({\bf Q},{\bf \Lambda})\}, 
\label{path1}\\  
S({\bf Q},{\bf \Lambda}) 
&=&\int_{0}^{t} dt'[-i 
{\bf \Lambda}\cdot {\dot {\bf Q}}\nonumber\\
&+&(1/2)\sum_{\alpha\beta}H_{\alpha\beta} 
({\bf Q}, \lambda_\alpha-\lambda_\beta)], 
\label{action1} 
\eea
\end{subequations}
where the vector ${\bf \Lambda}$ has components $\lambda_\alpha$:
node variables conjugated to the $Q_\alpha$ that impose charge conservation 
in the network. 

In the following, we define the functions $H_{\alpha\beta}$ as  the generating 
functions of the fast currents between nodes $\alpha$ and $\beta$. 
On the time scale 
$\Delta t\gg\tau_0$, the currents through isolated connectors are Markovian, 
so that all cumulants (irreducible correlators which 
are denoted by double angle brackets) of the transmitted charge 
$\langle\langle (Q_{\alpha\beta})^n\rangle\rangle$ are linear in $\Delta t$.
Following the standard notation in mesoscopic physics, 
\cite{Nazarov-Blanter} we define the current cumulants 
$\langle\langle (\tilde I_{\alpha\beta})^n\rangle\rangle$ as the coefficients
in 
\be
\langle\langle (Q_{\alpha\beta})^n\rangle\rangle=\Delta t
\langle\langle (\tilde I_{\alpha\beta})^n\rangle\rangle,
\label{I-cumulants}
\ee
where the tilde symbol has been introduced to distinguish the bare currents 
of each connector (the sources of noise) from the physical currents 
$I_{\alpha\beta}$ flowing through that same connector when it is placed 
into the network.
Then the generators $H_{\alpha\beta}$ are defined via the equation
\begin{equation} 
\langle\langle (\tilde{I}_{\alpha\beta})^n\rangle\rangle= 
\left.\frac{\partial^n H_{\alpha\beta}({\bf Q},\lambda_{\alpha\beta})} 
{(i\partial\lambda_{\alpha\beta})^n}\right|_{\lambda_{\alpha\beta}=0},
\label{noise3} 
\end{equation} 
and thus contain complete information about the statistics of the
noise sources. The $\lambda_{\alpha\beta}$ [eventually to be replaced
with with $\lambda_\alpha-\lambda_\beta$ in Eq.\ (\ref{pathint})] is
the generating variable for the current $\tilde{I}_{\alpha\beta}$.
The notion of current cumulants is useful because they are the time
independent objects, and thus have a time independent generators, Eq.\
(\ref{noise3}).  The generators $H_{\alpha\beta}({\bf
Q},\lambda_{\alpha\beta})$ depend in general on the full vector ${\bf
Q}$ and not just on the generalized charges of the neighboring nodes
$Q_\alpha$ and $Q_\beta$.  This may serve to incorporate long range
interactions between distant nodes.

 The charge $Q_{\alpha\beta}$ transfered through the connectors
[characterized by Eqs.\ (\ref{I-cumulants},\ref{noise3})] may be discrete.
However, the charge in the nodes $Q_\alpha$ is treated as an
effectively continuous variable in Eqs.\
(\ref{cons}-\ref{pathint}). This is justified if many charges in the
node participate in transport. Formally, this limit allows a
saddle-point evaluation of the propagator (\ref{path1}).

\subsection{Derivation of the Path Integral.}
\label{dpi}

To derive the path integral Eq.~(\ref{pathint}), we follow the usual procedure
\cite{kleinert} and first discretize time, $t= n \Delta t$ to derive an
expression for $U$ that is valid for propagation over one time step $\Delta t$.
Because of the separation of time scales $\tau_0 \ll \tau_C$, we can consider
$\Delta t$ as an intermediate time scale, 
\be \tau_0 \ll \Delta t \ll \tau_C\, .
\label{sep}
\ee 
The left inequality, $\tau_0 \ll \Delta t$, implies that the
transmitted charges $Q_{\alpha\beta}$ are Markovian. \cite{gardiner}
This means that charges transmitted in separate time intervals are
uncorrelated with each other.  While it is not necessary to specify
the source of the current correlation in the general formulation, it
is worth noting two examples.  In a mesoscopic point contact, the
correlation time $\tau_0$ has the interpretation of the time taken by
an electron wavepacket to pass the point contact.  In chemical
dynamics, it could be the time taken for a long molecule in solution
to traverse a filter.

In a time $\Delta t$, the probability that charge
$Q_{\alpha\beta}$ is transmitted between 
nodes $\alpha$ and $\beta$ can be written as the Fourier transform of the
exponential of a generating function $S_{\alpha\beta}$:
\be 
P_{\alpha\beta}(Q_{\alpha\beta},\Delta t) = 
\int \frac{d\lambda_{\alpha\beta}}{2\pi} 
\exp\{-i\lambda_{\alpha\beta} Q_{\alpha\beta}
+S_{\alpha\beta}(\lambda_{\alpha\beta})\}\, . 
\label{noise2} 
\ee 
The definition of the cumulant of transmitted charge is
\be
\langle\langle (Q_{\alpha\beta})^n\rangle\rangle=
\left.\frac{\partial^n S_{\alpha\beta}(\lambda_{\alpha\beta})} 
{(i\partial\lambda_{\alpha\beta})^n}\right|_{\lambda_{\alpha\beta}=0} .
\label{noise4} 
\end{equation} 
The Markovian assumption implies that the probability of transmitting charge
$Q_{\alpha\beta}$ in time $\Delta t$ followed by charge $Q'_{\alpha\beta}$ in
time $\Delta t'$ through any connector is given by the product of 
independent probability distributions.  This implies that the probability of
transmitting charge $Q_{\alpha\beta}$ in time $\Delta t+\Delta t'$ may be
calculated by finding all ways of independently transferring charge
$Q'_{\alpha\beta}$ in the first step and $Q_{\alpha\beta}-Q'_{\alpha\beta}$ in
the second step, 
\bea P(Q_{\alpha\beta},\Delta t+\Delta t') = \int
dQ'_{\alpha\beta} P(Q_{\alpha\beta}-Q'_{\alpha\beta},\Delta t')
\nonumber\\
\times
P(Q'_{\alpha\beta},\Delta t) \, ,
\label{con}
\eea 
which takes the form of a convolution of probabilities.  Applying a
Fourier transform to both sides of Eq.~(\ref{con}) with argument
$\lambda_{\alpha\beta}$ decouples the convolution into product of the two
Fourier transformed distributions.  Eq.~(\ref{noise2}) implies 
$S_{\alpha\beta}(\Delta t+\Delta t', \lambda_{\alpha\beta}) 
= S_{\alpha\beta}(\Delta t, \lambda_{\alpha\beta}) +
S_{\alpha\beta}(\Delta t', \lambda_{\alpha\beta})$.  
It then immediately follows that the generating function must be linear in time.  
Therefore, a time independent $H_{\alpha\beta}$
may be introduced: $S_{\alpha \beta} = \Delta t\, H_{\alpha \beta}$.  The
linear dependence of $S_{\alpha\beta}$ on time implies that all charge cumulants
(\ref{noise4}) will be proportional to time.  Therefore, we define the time
independent current cumulants, Eq.~(\ref{noise3}).

Different connectors are clearly uncorrelated for
$\Delta t \ll \tau_C$, which indicates that the total probability distribution
of transmitted charges is a product of the independent probabilities in each
connector:\cite{correlate}
\begin{equation} 
P[\{Q_{\alpha\beta}\}] = \prod_{\alpha>\beta}P_{\alpha\beta} 
[Q_{\alpha\beta},\Delta t]. 
\label{noise1} 
\end{equation} 
Thus far, the analysis is only valid for times much smaller than $\tau_C$. For
this case, the charges in the nodes will only slightly change.  
Since we wish to consider longer
times, we need to take into account the fact that charge transfer between
different nodes will be correlated as charge piles up inside the nodes.
This may be accounted for by imposing charge conservation Eq.~(\ref{cons})
during the time interval with a delta function, 
\bea
&&\delta(Q_{\alpha}-Q'_{\alpha}-\sum_{\beta}Q_{\alpha\beta})\nonumber\\
&&=\int\frac{d\lambda_{\alpha}}{2\pi}\exp\{-i\lambda_{\alpha}
[Q_{\alpha}-Q'_{\alpha}-\sum_{\beta}Q_{\alpha\beta}]\} \, .
\label{delta} 
\eea
Here, $Q'_{\alpha}$ is the charge in the node before the time interval while
$Q_{\alpha}$ is the charge accumulated in the node after the time interval is
over. In Eq.~(\ref{delta}), $\lambda_{\alpha}$ (referred to as a {\it counting
variable}) plays the role of a Lagrange multiplier.  The propagator is obtained
by multiplying the constraint (\ref{delta}) and the independent probability 
distribution (\ref{noise1}).  Representing the probabilities in their Fourier form
(\ref{noise2}) then yields
\begin{eqnarray}
&&{\tilde U}({\bf Q},{\bf Q}',Q_{\alpha\beta}, \Delta t) 
= \prod_{\alpha}\int \frac{d\lambda_{\alpha}}{2\pi} 
 \prod_{\alpha>\beta}\int \frac{d\lambda_{\alpha\beta}}{2\pi}\; 
\exp(S)\,,
\nonumber \\
&&S =
-i\sum_{\alpha}\lambda_{\alpha}(Q_{\alpha}-Q'_{\alpha}
-\sum_{\beta} Q_{\alpha\beta})
\nonumber\\
&&\qquad\qquad \;
+\sum_{\alpha>\beta}[- i \lambda_{\alpha\beta} Q_ {\alpha\beta} 
+ \Delta t H_{\alpha\beta}( {\bf Q}', \lambda_{\alpha\beta})]\, .
\label{path} 
\end{eqnarray}
{
The full propagator ${\tilde U}({\bf Q},{\bf Q}',Q_{\alpha\beta}, 
\Delta t)$ still keeps track of each individual connector contribution 
$Q_{\alpha\beta}$.} 
We now integrate out the fast 
fluctuations to obtain the dynamics of the slow variables.  
This may be done by using the identity
$\sum_{\alpha} \lambda_\alpha \sum_{\beta} Q_{\alpha\beta} =
\sum_{\alpha>\beta} ( \lambda_\alpha Q_{\alpha\beta}+\lambda_\beta
Q_{\beta\alpha} )$ and $Q_{\alpha\beta}=-Q_{\beta\alpha}$.  The integration
over $Q_{\alpha\beta}$ gives a delta function of argument
$\lambda_{\alpha\beta}- (\lambda_{\alpha}-\lambda_{\beta})$, so that the
$\lambda_{\alpha\beta}$ integrals may be trivially done. We obtain
\bea 
U({\bf Q},{\bf Q}',\Delta t)
&=& \prod_{\alpha}\int
\frac{d\lambda_{\alpha}}{2\pi}
\exp \big\{-i\sum_{\alpha}\lambda_{\alpha}(Q_{\alpha}-Q'_{\alpha}) 
\nonumber\\
&&\quad
+\Delta t\sum_{\alpha>\beta}H_{\alpha\beta}
({\bf Q}',\lambda_{\alpha}-\lambda_{\beta})\big\}.
\label{p1} 
\eea
This is the general result for the one step propagator.  If any two
nodes are unconnected, $H_{\alpha\beta}$ is zero.  

An important comment is in order: because $H_{\alpha\beta}$ changes 
slightly over
the time period, which in turn affects the probability of transmitting
charge through the contacts, it is not clear at what part of the time
step $H_{\alpha\beta}$ should be evaluated.  This ambiguity exists
because  
our theory is not microscopic.  Rather, it takes the
microscopic noise generators as an input.  This ambiguity gives the
freedom of stochastic quantization.\cite{oporder} The same problem
also occurs in quantum mechanical path integrals, and its source there
is an ambiguity in operator ordering.\cite{Langevin} As we are
interested in the large transporting charge limit, $\gamma \gg 1$,
and evaluate the integrals in leading order saddle-point
approximation, this ambiguity will not affect the results. \cite{us} 
For calculations beyond the large transporting charge limit, the canonical
variables ${\bf Q}$ and ${\bf \Lambda}$ need to be properly ordered,
which can only be done with a microscopic theory.  For example, the
master equation discretized in time as discussed in Sec.\ \ref{meq}
requires the placement of ${\bf \Lambda}$ operators in front of ${\bf Q}$
operators, since the generating functions $H_{\alpha\beta}$ of the
transition probabilities depend on the state of the system at the
beginning of the time period.

To extend the propagator (\ref{p1}) to longer times $t= n \Delta t$, we use 
the composition
property of the evolution operator (also known as the Chapman-Kolmogorov
equation\cite{gardiner}). This requires separate $\{Q_\alpha\}$ integrals at
each time step, so that for $n$ time steps there will be $n-1$ integrals
over ${\bf Q}$, while each of the $n$ one-step propagators comes with its own
${\bf \Lambda}$ integral, ${\bf \Lambda} = \{\lambda_\alpha\}$.  Inserting our
expression for the $\Delta t$ step propagator Eq.~(\ref{p1}), we find 
\begin{subequations}
\begin{widetext}
\be
U({\bf Q}_f,{\bf Q}_i,t) =\int d{\bf \Lambda}_0\prod_{k=1}^{n-1}
\int d{\bf Q}_k\, d{\bf \Lambda}_k\exp \left[ \sum_{k=0}^{n-1}
-i {\bf \Lambda}_k\cdot \left ( {\bf Q}_{k+1}-{\bf Q}_{k} \right) + \Delta t
H({\bf Q}_{k},{\bf \Lambda}_k)\right],
\label{path2} 
\ee
\end{widetext}
with
\be
H({\bf Q}_k,{\bf \Lambda}_k) =
\sum_{\alpha>\beta}H_{\alpha\beta} 
\left[Q_{\alpha,k}; \lambda_{\alpha,k}-\lambda_{\beta,k}\right] \, ,
\label{action2} 
\ee 
\end{subequations}
where we have introduced the notations $d{\bf Q}_k=\prod_{\alpha}dQ_{\alpha,k}$
and $d{\bf \Lambda}_k=\prod_{\alpha}(d\lambda_{\alpha,k}/2\pi)$.
We are now in a position to take the continuous time limit.  Writing ${\bf
Q}_{k+1}-{\bf Q}_k=\Delta t\, {\dot {\bf Q}}$, which is valid because the
charge in any node changes only slightly over the time scale $\Delta t$, the
action of this discrete path integral has the form $S = \Delta t \sum_{k=1}^n
S_k$, which goes over into a time integral in the continuous limit. Using the
standard path integral notation $\int {\cal D}{\bf Q}{\cal D}{\bf \Lambda} = \int
d{\bf \Lambda}_0\prod_{k=1}^{n-1} \int d{\bf Q}_kd{\bf
\Lambda}_k$, and invoking the symmetry
$H_{\alpha\beta}(\lambda_\alpha-\lambda_\beta)=
H_{\beta\alpha}(\lambda_\beta-\lambda_\alpha)$ we recover Eq.~(\ref{pathint}).
The only explicit constraint on the path integral comes with the charge
configurations at the start and finish, ${\bf Q}_i$ and ${\bf Q}_f$.  We also
note that $H_{\alpha\beta}$ depends on any external parameters such as
voltages or chemical potentials driving the charge ${\bf Q}$.

In the simplest case of one charge and counting variable, the form of
the path integral is the same as the (Euclidian time) path integral
representation of a quantum mechanical propagator in phase space with
position coordinate $Q$ and momentum coordinate
$\lambda$.\cite{Langevin} The differences with the quantum version are
that the propagator evolves probability distributions, not amplitudes
(similarly to Ref.\ \onlinecite{cardy}), as well as the fact that the
``Hamiltonian'' $H=(1/2)\sum_{\alpha\beta} H_{\alpha\beta}({\bf Q},
\lambda_\alpha-\lambda_\beta)$ is not really a Hamiltonian, but rather
a current cumulant generating function and therefore is not Hermitian
in general. Even so, because of the similarity we shall refer to $H$
as the Hamiltonian from now on.

\subsection{Absorbed Charges, Boundary Conditions and Correlation Functions.}
\label{ac}

A useful special case occurs when one has absorbed charges.  These are charges
that vanish into (or are injected from) 
absorbing nodes without altering the system dynamics.  In
mesoscopics for example, the absorbing nodes are metallic reservoirs.
Formally, we divide the charges into those that are conserved and those that
are absorbed: ${\bf Q}= \{ {\bf Q}^c, {\bf Q}^a\}$, where the subset of
absorbed charges ${\bf Q}^a = \{Q^a_\alpha\}$ does not appear in
$H_{\alpha\beta}$.  We do the same for the corresponding counting variables:
${\bf \Lambda}= \{ {\bf \Lambda}^c, {\bf \Lambda}^a\}$.  Because
$H_{\alpha\beta}$ does not depend on ${\bf Q}^a$, these charges may be
integrated out by integrating the action by parts,
\bea
i\int_{0}^{t} dt'\; {\bf \Lambda}^a\cdot {\dot {\bf Q}^a}=
&-& i\int_{0}^{t} dt'\; {\bf Q}^a\cdot {\dot {\bf \Lambda}^a}
\nonumber\\
&+& i\, ({\bf \Lambda}^a_f\cdot {\bf Q}^a_f
-{\bf \Lambda}^a_i\cdot {\bf Q}^a_i)\, ,
\label{ibp}
\eea
and then functionally integrating over ${\bf Q}^a$ to obtain ${\bf \delta}
({\dot {\bf \Lambda}}^a)$, where ${\bf \delta}$ is a functional delta
function.  This immediately constrains the ${\bf \Lambda}^a$ to be constants
of motion so the functional integration over ${\bf \Lambda}^a$ becomes a
normal integration, ${\cal D}{\bf \Lambda}^a \rightarrow d{\bf\Lambda}^a$.  The absorbed kinetic terms in the action may then be integrated
to obtain
\begin{subequations}
\label{pathint1}
\begin{equation} 
U({\bf Q}_f,{\bf Q}_i,t) 
=\int d{\bf \Lambda}^a
\int\! {\cal D}{\bf Q}^c{\cal D}{\bf \Lambda}^c
\exp\left\{S({\bf Q}, {\bf \Lambda})\right\}, \,
\label{path3}
\end{equation}
\bea
&&\!\!\!S({\bf Q}, {\bf \Lambda}) 
=\int_{0}^{t} dt'[-i 
{\bf \Lambda}^c\cdot {\dot {\bf Q}}^c+
(1/2)\sum_{\alpha\beta}H_{\alpha\beta} 
(\lambda_{\alpha}-\lambda_{\beta})] 
\nonumber\\
&&\qquad\qquad\qquad\qquad\qquad
-i {\bf \Lambda}^a\cdot({\bf Q}^a_f-{\bf Q}^a_i). 
\label{action3} 
\eea
\end{subequations}

Often one is interested in the probability to transmit some amount of charge
through each of the absorbing nodes. By applying a Fourier transform to
Eq.~(\ref{path3}) with respect to ${\bf Q}^a(t)-{\bf Q}^a(0)$ we remove the
last term in Eq.\ (\ref{action3}) and obtain the path integral representation
for the characteristic function $Z$ which generates current moments at every
absorbing node
\begin{subequations}
\label{pathint2}
\bea
Z({\bf \Lambda}^a) 
&=&
\int\! {\cal D}{\bf Q}^c{\cal D}{\bf \Lambda}^c
\exp\{S({\bf Q},{\bf \Lambda})\}, 
\label{path4} \\
S({\bf Q}, {\bf \Lambda}) 
&=&\int_{0}^{\,t}\!\! dt'[-i 
{\bf \Lambda}^c\cdot {\dot {\bf Q}}^c
\nonumber\\
&&\qquad+(1/2)\sum_{\alpha\beta}H_{\alpha\beta} 
(\lambda_\alpha-\lambda_\beta)]  \, .
\label{action4} 
\eea 
\end{subequations}
Note that the counting variables ${\bf \Lambda}^a$ enter the action
(\ref{action4}) only as a set of constant parameters.  The initial condition
in the path integral (\ref{pathint2}) is given by the initial charge states ${\bf
Q}^c(0)$.  There is a choice of the final condition: by fixing the final ${\bf
Q}^c(t)$ one obtains the distribution of the conserved charge subject to this
constraint, while by fixing ${\bf \Lambda}^c(t)$ the corresponding
characteristic function is obtained.  The choice of ${\bf \Lambda}^c(t)=0$ in
Eq.\ (\ref{pathint2}) gives the characteristic function of the absorbed charge
under the condition that the conserved charge is not being monitored, i.e.\
the final charge state is integrated over.  Therefore $\ln Z$ becomes the
generator of the FCS, defining the charge cumulants at the absorbing node, 
\be
\langle \langle [{Q}^a_\alpha(t)-{Q}^a_\alpha(0)]^n \rangle \rangle =
\left. \frac{\partial^n \ln Z} {\partial (i \lambda^a_\alpha)^n}\right|_{{\bf
\Lambda}^a=0}  .
\label{gen2}
\ee
In the long time limit, this quantity is proportional to time, independent of
the details of the boundary conditions.

Alternatively, in the short time limit one may calculate irreducible
correlation functions of absorbed and conserved current fluctuations,
${\bf I}=\dot {\bf Q}$.  These correlation functions can be obtained by
extending the time integral in (\ref{action1}) to infinity, introducing
sources\cite{Langevin} in the action, $S\rightarrow S+i \int dt\, {\bf
\chi}(t)\cdot{\bf I}(t)$, and applying functional derivatives with respect to
${\bf \chi}$.  Repeating the steps leading to Eqs.\ (\ref{pathint2}), 
we find that variables $\lambda_{\alpha}$ in the Hamiltonian
in Eq.\ (\ref{action4}) have to be shifted
$\lambda_{\alpha}\to\lambda_{\alpha}+\chi_{\alpha}$.  Then, the irreducible
current correlation function is given by 
\be \langle \langle I_{\alpha_1}(t_1)
\cdots I_{\alpha_n}(t_n)\rangle\rangle =\left.  \frac{\delta^n\ln Z[{\bf
\chi}]}{\delta i\chi_{\alpha_1}(t_1) \cdots\delta i\chi_{\alpha_n}(t_n)}
\right|_{\,{\bf \chi}=0}  .
\label{corr}
\ee 
With these correlation functions, one may calculate 
for example the frequency dependence of current cumulants.\cite{freqdep}

\subsection{The Saddle Point Approximation.}
\label{sp}

If the Hamiltonian has some dimensionless large prefactor, then the path
integral (\ref{pathint}) may be evaluated using the saddle point
approximation, which is justified below.  At the saddle point,
(where the first variation of the action vanishes), we can write equations of
motion analogous to the Hamiltonian equations of classical mechanics:
\be
i {\dot {\bf Q}}^c = \frac{\partial}{\partial {\bf \Lambda}^c }
H({\bf Q}^c, {\bf \Lambda}), \quad
i {\dot {\bf \Lambda}}^c = -\frac{\partial}{\partial {\bf Q}^c }
H({\bf Q}^c, {\bf \Lambda}) ,
\label{eom0}
\ee
where $H({\bf Q}^c, {\bf \Lambda})=(1/2)\sum_{\alpha\beta}
H_{\alpha\beta}({\bf Q}^c;\lambda_\alpha-\lambda_\beta)$.  There may
be many saddle point solutions in general, and one has to sum over all
of them.  Eqs.~(\ref{eom0}) are solved subject to the temporal
boundary conditions and generally describe the relaxation of the
conserved charges from the initial state to a stationary state $\{
{\bf \bar Q}^c , {\bf \bar \Lambda}^c \}$ on a time scale given by
$\tau_C$, the dynamical time scale of the nodes. These stationary
coordinates are functions of any external parameters as well as the
(constant) absorbed counting variables ${\bf \Lambda}^a$.  In the
saddle point approximation, the action takes the form $S=S_{sp} +
S_{fluc}$.\cite{us} The term $S_{sp}$ is the contribution to the
action from the solution of the equations (\ref{eom0}), which
describes the evolution of the system from the initial to the final
state.  The term $S_{fluc}$ describes fluctuations around the saddle
point and is suppressed compared to the saddle-point contribution, if
the Hamiltonian has a large prefactor (in analogy to the
$\hbar$-expansion of quantum mechanics).  Physically, the validity
condition for the saddle point approximation is that there should be
many (transporting) charge carriers in the nodes.  For times longer
than the charge relaxation time of the node, the dominant contribution
is from the stationary state only, where the saddle-point part of the
action is simply linear in time:
\be S_{sp}( {\bf \bar Q}, {\bf \bar \Lambda}) = t
H({ \bf \bar Q}, {\bf \bar \Lambda} ), \qquad t \gg \tau_C\, .
\label{gen}
\ee 
The linear time dependence of Eq.~(\ref{gen}) indicates that the dynamics are
Markovian on a long time scale.  It is the fact that the contribution $S_{sp}$
emerges in a dominant way which makes the approach given here a powerful tool
to analyze the counting statistics of transmitted charge.

We now discuss the large parameter that justifies the saddle point
approximation.  The boundary conditions on the charge in the absorbing nodes
fix a (dimensionless) charge scale of the system, $\gamma$.  All charges in
the network are scaled accordingly, ${\bf Q} \rightarrow \gamma {\bf Q}$.  We
make the assumption that there is a one parameter scaling of the Hamiltonian,
$H\rightarrow \gamma H$.  The time is also scaled by $\tau_C$, the time scale of
charge relaxation in the nodes. The dimensionless action is now $S=\gamma
\int_0^{t/\tau_C} dt' (-i{\dot Q}\lambda +\tau_C H)$. The saddle point action
is proportional to $\gamma t/\tau_C$, while the fluctuation contribution will
be of order $t/\tau_C$.  We note that the parameter $\gamma$ is related to
(though not necessarily the same as) the separation of time scales,
$\tau_C/\tau_0$, needed to derive the path integral.  For the mesoscopic
conductors considered in the example section \ref{cavityex} of this paper, the
charge scale is set by the maximum number of semiclassical states on the
cavity involved in transport, $\gamma=\Delta\mu N_F\gg 1$, the bias times the
density of states at the Fermi level.  On the other hand, for the chaotic
cavity, $\tau_C/\tau_0=\gamma/(G_L+G_R)$, where $G_{L,R}\gg 1$ are the
dimensionless conductances of the left and right point contact.

\subsection{Relation to the master Equation and  Doi-Peliti Technique.}
\label{meq}

The evolution operator $U({\bf Q}, {\bf Q}', t)$ may be interpreted as a
Green function of a differential equation which determines the propagation in
time of an initial probability distribution $\Gamma({\bf Q})$.  In the theory
of stochastic processes, such a differential equation is called a master
equation.  A natural question that arises is the relationship of the formalism
presented here to other approaches to stochastic problems.

The most general type of Markovian master equation for discrete states and
discrete time is of the form
\be
\Gamma_n(t_{k+1}) = \sum_m P_{nm}(t_{k+1},t_k) \Gamma_m (t_k)\, ,
\label{me}
\ee
{
where $\Gamma_m(t_k)$ is the probability  to be in state $m$ 
at time $t_k$ and
$P_{nm}$ is the transition probability from state $m$ to state $n$.  
The state is described by a vector $n = (n_1,\dots,n_N)$ whose components 
are the charges $n_\alpha$ of each node $\alpha$.}
The
Markovian assumption implies that $t_{k+1}-t_k=\Delta t$ is greater than the
correlation time, $\tau_0$.  If we further assume that the probability to make
a transition to another state is small, $P_{nm} \ll 1$ for $n \ne m$, so that
the transition probability is only linear in $\Delta t$, a transition rate
$W_{nm} = P_{nm}/\Delta t$ may be defined.  It then follows that we may write
a differential master equation,
\be
{\dot \Gamma}_n(t) = \sum_m [W_{nm} \Gamma_m(t) - W_{mn} \Gamma_n(t)]\, .
\label{cme}
\ee
Eq.~(\ref{cme}) is the starting point for the Doi-Peliti technique, \cite{dp}
where one formally maps the space of physical states to the {
Fock space of
states $\vert n\rangle = (a_1^\dagger)^{n_1} 
\dots (a_N^\dagger)^{n_N} 
\vert 0 \rangle$}, where $n$ is the
number of charges.  The entire state of the system is {
expressed}
 by a vector
$\vert \Psi\rangle = \sum_n \Gamma_n \vert n\rangle$ which weights the states
$\vert n\rangle$ with their probabilities $\Gamma_n$.  Thus, the master equation
(\ref{cme}) may be interpreted as a many-body Schr\"odinger equation where the
rates $W_{mn}$ are incorporated into a Hamiltonian in a second-quantized form.
One may then write a coherent-state path integral over the variables $a$, and
$a^\dagger$ for this many-body quantum system and perform perturbation
expansions along with the renormalization group.\cite{cardy} This procedure
eventually involves taking the continuum limit so the discrete charge states
become continuous.

Let us now consider how our formalism is related to the master equation or the
Doi-Peliti technique.  According to the results of Sec.\ \ref{gfor}, our
stochastic path integral, Eq.(\ref{pathint}) solves the continuum variable
version of Eq.~(\ref{me}) with the transition probabilities given by the one
step propagator $U({\bf Q},{\bf Q}', \Delta t)$.
{\em In general, the transition probabilities are neither small nor linear in
time for $\Delta t>\tau_0$}.  It is instructive nevertheless to consider the
special case of processes where $H\tau_0\ll 1$, when we can expand the one-step
propagator (\ref{p1}) to first order in $\Delta t$,
\bea 
U({\bf Q},{\bf Q}', \Delta t) 
&\approx &\delta({\bf Q}-{\bf Q}') 
\nonumber\\
&+&\Delta t
\int d{\bf \Lambda}e^{-i {\bf \Lambda} \cdot ({\bf Q}-{\bf Q}')} H({\bf Q}',
{\bf \Lambda}).\quad
\label{expand}
\eea
Defining the Fourier transform of the generating function as ${\tilde H}({\bf
Q},{\bf Q}')$, the differential equation governing the evolution of a
probability distribution of charges $\Gamma({\bf Q})$ is then
\be
{\dot \Gamma}({\bf Q}, t) = 
 \int d{\bf Q}'\; {\tilde H}({\bf Q},{\bf Q}') \Gamma({\bf Q}', t) .
\label{master}
\ee
Comparison with the continuous version of the master equation (\ref{cme}),
\be
{\dot \Gamma}({\bf Q}, t) = 
\int d{\bf Q}'\, [W({\bf Q}, {\bf Q}') \Gamma({\bf Q}', t) - 
 W({\bf Q}', {\bf Q})\Gamma({\bf Q}, t)] ,
\label{ccme}
\ee
indicates that ${\tilde H}$ is related
to $W$.  The Hamiltonian may be expressed in terms of the transition
kernel\cite{lax} as,
\be
H({\bf Q}',{\bf \Lambda}) = \int d{\bf Q} 
\left[e^{i ({\bf Q}-{\bf Q}')\cdot {\bf \Lambda}} -1\right] W({\bf Q},
     {\bf Q}') ,
\label{translation}
\ee 
where the normalization of probability is expressed by $H({\bf
Q}',0)=0$.  {
Eq.\ (\ref{translation}) is an important result,
because it allows the conversion of the master equation (\ref{ccme})
into the stochastic path integral (\ref{pathint}).}

We would like to stress that our formalism is not simply equivalent to
the differential master equation (\ref{ccme}) (and therefore the
Doi-Peliti technique), but that it allows the treatment of a {
complementary} class of problems.  
Our formalism assumes
effectively continuous charge, and thus cannot resolve effects due to the
discreteness of charge on the nodes. Such effects are present in the
master equation\ (\ref{cme}).  In contrast, the differential master
equation assumption, $H \tau_0 \ll 1$ (which simply states that
transition probabilities are small in the time interval $\tau_0$)
is not required.  Our formalism is especially important when this is not the
case, ${i.e.}\ H \tau_0 \sim 1$.

This is illustrated by the simple example from mesoscopics of two metallic
reservoirs connected by a single electron barrier with hopping probability $p$
and bias $\Delta\mu$ at zero temperature.  For a time interval $\Delta t$
larger than the correlation time $\tau_0=\hbar/\Delta\mu$ (the time scale for
an electron wavepacket to transverse the barrier), $\Delta t/\tau_0$ electrons
approach the barrier and either are transmitted or reflected.  Mathematically,
this is a classical binomial process with the generator
\be
S =  (\Delta t/\tau_0 )\ln [1 + p (e^{i e \lambda}-1)] .
\label{pc}
\ee
As this action is the starting point of many mesoscopic implementations of the
formalism, it is an important example.  Since the action is proportional to
the large parameter $\Delta t /\tau_0>1$, for $p\sim 1$ the expansion of
$\exp(S)$ to first order in $\Delta t$ is strictly forbidden, effectively not
allowing a first order differential master equation.  Only in the limit $p \ll
1$, ($i.e.$ when Eq.~(\ref{pc}) describes a Poissonian process) may the
logarithm be expanded to first order.  This suggests that Eq.~(\ref{ccme})
describes the slow dynamics of systems whose fast transitions are Poissonian
in nature.  A more general type of dynamics such as the binomial distribution
may only be found using the continuous charge state master equation in
discrete time (\ref{me}).

\section{The Field Theory}
\label{ft}

From the stochastic network, Fig.\ \ref{network}, it is straightforward to go
to spatially continuous systems as the spacing between the nodes is taken to
zero.  The goal is to introduce a Hamiltonian functional $ h (\rho, \lambda)$
whose arguments are the charge density $\rho$ and the counting field functions
$\lambda$, that are themselves functions of space and time. We may then
replace $(1/2)\sum_{\alpha,\beta}H_{\alpha,\beta} \rightarrow \int dz h (\rho,
\lambda)$.  Our description is local, so in the model each node is only
connected to its nearest neighbors.  We first derive the one dimensional field
theory with one charge species in detail, and then generalize to multiple
dimensions and charge species.
\begin{figure}[t]
\centerline{\includegraphics[width=7cm]{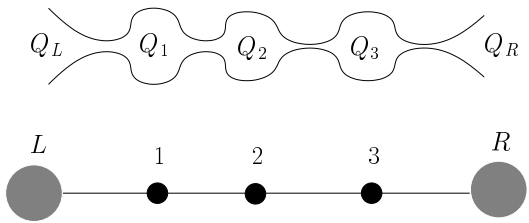}}
\caption{A one dimensional lattice of nodes connected on both ends
to absorbing reservoirs.  This situation could represent a series of
mesoscopic chaotic cavities connected by quantum point contacts.}
\label{lattice}
\end{figure}

Consider a series of identical, equidistant nodes separated by a distance
$\Delta z$.  This nodal chain could represent a chain of chaotic cavities,
Fig.\ \ref{lattice}, in a mesoscopic context.\cite{Oberholzer1,Ob2} The sum
over $\alpha$ and $\beta$ becomes a sum over each node in space connected to
its neighbors.  The action for this arrangement is
\be
S = \int_0^t dt' \sum_{\alpha} \{-\lambda_\alpha {\dot Q}_\alpha
+H(Q_{\alpha}, Q_{\alpha-1};\lambda_{\alpha}-\lambda_{\alpha-1})\}\; ,
\label{act1}
\ee
where for simplicity we have chosen real counting variables,
$i\lambda_\alpha\rightarrow \lambda_\alpha$.  The imaginary counting variables
will be restored at the end of the section.  The only constraint made on $H$
is that probability is conserved, $H(\lambda_{\alpha}-\lambda_{\alpha-1})=0$
for $\lambda_{\alpha}=\lambda_{\alpha-1}$.  We now derive a lattice field
theory by formally expanding $H$ in $\lambda_{\alpha}-\lambda_{\alpha-1}$ and
$Q_{\alpha}-Q_{\alpha-1}$.  Only differences of the counting variables will
appear in the series expansion, while we must keep the full $Q$ dependence of
the Hamiltonian.  If there are $N \gg 1$ nodes in the lattice, for fixed
boundary conditions the difference between adjacent variables,
$\lambda_{\alpha}-\lambda_{\alpha-1}$ and $Q_{\alpha}-Q_{\alpha-1}$ will be of
order $1/N$, and therefore provides a good expansion parameter.  The expansion
of the Hamiltonian (\ref{act1}) to second order in the difference variables
gives
%
\bea
H &=&  \frac{\partial H}{\partial
 \lambda_{\alpha}}\, (\lambda_{\alpha}-\lambda_{\alpha-1}) 
 + \frac{1}{2}\,\frac{\partial^2 H}{\partial
 \lambda_{\alpha}^2} (\lambda_{\alpha}-\lambda_{\alpha-1})^2 
 \nonumber\\
&+& 
\frac{\partial^2 H}{\partial
    Q_{\alpha}\partial \lambda_{\alpha}} (Q_{\alpha}-Q_{\alpha-1})
  (\lambda_{\alpha}-\lambda_{\alpha-1})\, , 
\label{expandh}
\eea
%
where the expansion coefficients are evaluated at
$\lambda_{\alpha}=\lambda_{\alpha-1}$ and $Q_{\alpha}=Q_{\alpha-1}$ and are
functions of $Q_{\alpha-1}$.  Terms involving only differences of
$Q_\alpha-Q_{\alpha-1}$ are zero because
$H(\lambda_{\alpha}-\lambda_{\alpha-1})=0$ for
$\lambda_{\alpha}=\lambda_{\alpha-1}$.  All terms in Eq.~(\ref{expandh}) need
explanation.  First, the expression ${\partial H}/{\partial\lambda_\alpha}$ is
the local current at zero bias (because the charges in adjacent nodes are
equal) which will usually be zero.  There may be circumstances
where this term should be kept,\cite{sbs} but we do not consider them here.
The term $ \partial^2 H/{\partial Q_{\alpha} \partial\lambda_{\alpha}} =
-G(Q_{\alpha-1})$ is the linear response of the current to a charge
difference.  Hence, $G$ is the generalized conductance\cite{cap} of the
connector between nodes $\alpha$ and $\alpha-1$.  $\partial^2 H/{\partial
\lambda_{\alpha}^2} = C(Q_{\alpha-1})$ is the current noise through the same
connector because $H$ is the generator of current cumulants.

We are now in a position to take the continuum limit by replacing the node
index $\alpha$ with a coordinate $z$, introducing the fields $Q(z),
\lambda(z)$, and making the expansions
\begin{subequations}
\begin{eqnarray}
\lambda_{\alpha}-\lambda_{\alpha-1}\!\to\! \lambda'\Delta z+
(1/2)\lambda''(\Delta z)^2+{\cal O}(\Delta z)^3,\quad 
\label{expa}\\
Q_{\alpha}- Q_{\alpha-1}
\!\to\! Q'\Delta z+
(1/2) Q''(\Delta z)^2+{\cal O}(\Delta z)^3.\quad
\label{expb}
\end{eqnarray}
\end{subequations}
The action may now be written in terms of intensive fields
by scaling away $\Delta z$,
\bea
&&H\rightarrow   h (\rho,\lambda) \Delta z,\quad 
Q_\alpha \rightarrow \rho(z) \Delta z,\nonumber\\ 
&&G_\alpha (\Delta z)^2 \rightarrow D(\rho),\quad 
C_\alpha \Delta z \rightarrow F(\rho)\, ,
\label{redef}
\eea
and taking the limit $\sum_\alpha H \rightarrow \int dz h (\rho, \lambda)$.
One may check that expanding the Hamiltonian to higher than second order in
$\Delta z$ will result in terms suppressed by powers of $\Delta z/L$ and
consequently vanish as $\Delta z\rightarrow 0$.  This scaling argument
for the field theory is analogous to Van Kampen's 
size expansion.\cite{vankampen}  Though the lattice spacing
$\Delta z$ does not appear in the continuum limit, it provides a physical
cut-off for any ultra-violet divergences that might appear in a loop
expansion.

These considerations leave the one dimensional action as
\be
S = -\int_0^t dt'\int_0^L dz \left [\lambda  {\dot \rho}+  D\, \rho' \lambda'
- \frac{1}{2}F\, (\lambda')^2 
\right ] .
\label{act1d}
\ee
Here $D$ is the local diffusion constant and $F$ is the local noise density
which are discussed in detail below.  It is very important that these two
functionals $D,F$ are all that is needed to calculate current statistics.
Classical field equations may be obtained by taking functional derivatives of
the action with respect to the charge and counting fields: $\delta S /\delta
\rho(z) = \delta S /\delta \lambda(z) =0$ to obtain the equations of motion,
\be
{\dot \lambda} =-\frac{1}{2}\frac{\delta F}{\delta \rho} \,(\lambda')^2
- D\lambda'' , \qquad
  {\dot \rho} = [-F  \lambda' + D \rho']' .
\label{eom1}
\ee
From the charge equation, one can see immediately that the term inside the
derivative may be interpreted as a current density so that local charge
conservation is guaranteed.  We have to solve these coupled differential
equations subject to the boundary conditions
\bea
\rho(t, 0) =  \rho_L(t), \quad \rho(t, L) =  \rho_R(t),\nonumber\\
 \lambda(t, 0) =  \lambda_L(t), \quad \lambda(t, L) =  \lambda_R(t),
\label{bcs1}
\eea
where $\rho_L(t)$, $\rho_R(t)$, $\lambda_L(t)$, and $\lambda_R(t)$ are
arbitrary time dependent functions.  Functions $\rho_L(t)$ and $\rho_R(t)$ are
the charge densities at the far left and right end of the system which may be
externally controlled. Functions $\lambda_L(t)$ and $\lambda_R(t)$ are the
counting variables of the absorbed charges at the far left and right end which
count the current that passes them.

Once Eqs.~(\ref{eom1}) are solved subject to the boundary conditions
(\ref{bcs1}), the solutions $\rho(z,t)$ and $\lambda(z,t)$ should be
substituted back into the action (\ref{act1d}) and integrated over time and
space.  The resulting function,
$S_{sp}[\rho_L(t),\rho_R(t),\lambda_L(t),\lambda_R(t),t,L]$ is the generating
function for time-dependent cumulants of the current distribution.  Often, the
relevant experimental quantities are the stationary cumulants.  These are
given by neglecting the time dependence, finding static solutions, ${\dot
\rho} ={\dot \lambda}=0$, and imposing static boundary conditions.  Similarly
to section \ref{meq}, we can also introduce sources $\int dtdz\,
\chi(z,t)\rho(z,t)$ and calculate density correlation functions.

To estimate the contribution of the fluctuations to the action, it is useful
to define dimensionless variables.  The boundary conditions $\rho_L$, and
$\rho_R$ provide the charge density scale $\rho_0$ in the problem, so we
define $\rho(z) =\rho_0 f(z)$, where $f\sim 1$ is an occupation.  We
furthermore rescale $z\to Lz$, and $t\to\tau_D t$, where $\tau_D=L^2/D$ is the
diffusion time, thus obtaining
\be
S= -L \rho_0\int_0^{\,t}\! dt'\!\int_0^1 dz' 
\left [\lambda  {\dot f}+  f' \lambda'-\frac{F}{2D\rho_0}(\lambda')^2 
\right ] .
\label{act1d2}
\ee
We assume that the combination ${F}/{D\rho_0}$ is of order 1.  From
Eq.~(\ref{act1d2}), the dimensionless large parameter is $\gamma=\rho_0L\gg
1$, i.e.\ the number of transporting charge carriers.  As in Sec.\ \ref{sp},
the saddle point contribution is of order $\gamma t/\tau_D$, while the
fluctuation contribution is of order $t/\tau_D$.

Repeating this derivation in multiple dimensions with $N$ charge species
$\rho=\{\rho_i({\bf r}) \}$ and counting fields $\Lambda =\{\lambda_i({\bf
r})\}$, $i=1,\ldots, N$ yields the action
\be
S = -\int_0^{\,t} \! dt'\! \int_\Omega d{\bf r} \; [\,  {\Lambda}
{\dot \rho} + \nabla{\Lambda} \, {\hat D}\, \nabla{\rho} -
(1/2)\nabla{\Lambda}\, {\hat F}\, \nabla{\Lambda}
\, ] ,
\label{fieldaction}
\ee
where tensor notation is used and we have introduced ${\hat
F}_{ij}=\partial_{\lambda_i}\partial_{\lambda_j} h $ and ${\hat
D}_{ij}=-\partial_{\rho_i} \partial_{\lambda_j} h $ as general matrix
functionals of the field vector ${\rho}$ and coordinate ${\bf r}$ which should
be interpreted as noise and diffusion matrices. If the medium is isotropic,
then the vector gradients simply form a dot product.  It should be emphasized
that the vectors appearing are vectors of different species of charge fields,
as all node delimitation has been accounted for in the spatial integration.
The functional integral now runs over all field configurations that obey
the imposed boundary conditions at the surface $\partial \Omega$.  
Classical field equations may be formally obtained by taking functional
derivatives of the action with respect to the charge and counting
fields as in the 1D case.

As in any field theory, symmetries of the action play an important
role because they lead to conserved quantities.  We first note that
the Hamiltonian $h(\rho, \nabla \rho, \nabla { \Lambda})$ is a
functional of $\nabla {\Lambda}$ alone with no ${\Lambda}$ dependence.
This symmetry is analogous to gauge invariance, and leads to the
equation of motion 
\be
{\dot \rho} + \nabla \cdot {\bf j}=0\, ,\quad
{\bf j}=-{\hat D}\nabla \rho + {\hat F}\nabla {\Lambda}\, ,
\label{conslaw1}
\ee  
which can
be interpreted as conservation of the conditional current ${\bf j}$. 
The next symmetry is
related to the invariance under a shift in the space and time
coordinates $\{{\bf \delta \bf r}, \delta t\}$. This symmetry leads to
equations analogous to the conservation of the local energy/momentum
tensor.\cite{landau} We do not explicitly give this quantity because
it is rather cumbersome in the general case. However, for the
stationary limit (where $\dot \rho$ and $\dot \lambda$ vanish) and for
symmetric diffusion and noise tensors, the one charge species
conservation law is relatively simple and is given by
\begin{subequations}
\label{conslaw2}
\bea
&&\sum_m \nabla_m T_{mn} = 0\, ,\quad \\ 
&&T_{mn} = j_m (\nabla_n \lambda) - (\nabla_n \rho)\, (\hat D \nabla \lambda)_m -
h\, \delta_{mn} \, .\quad
\eea
\end{subequations}
For the special case of a one dimensional geometry, 
 the Hamiltonian itself is the conserved quantity (see Sec.~\ref{wire}).

In the continuum limit, all terms of higher order in $\Lambda$ are suppressed
so that the action is quadratic in the $\Lambda$ variables.
This fact may be viewed as a consequence of the central limit theorem and
confirms the observation made by Nagaev that local noise in the mesoscopic
diffusive wire (see Sec.\ \ref{wire}) is Gaussian.\cite{Nagaev1} To further
clarify the physical meaning of $D$ and $F$, and also to make connection with
previous work,\cite{Langevin} we restore the complex variables, $\Lambda\to
i\Lambda$, and make a Hubbard-Stratronovich transformation by introducing an
auxiliary vector field ${\bf \nu}$,
\bea
&&\exp\{-(1/2)\nabla\Lambda\, {\hat F}\, \nabla\Lambda\} 
\nonumber\\
&&
=(\det {\hat F})^{-\frac{1}{2}}
\int {\cal D} {\nu} \,
\exp\{-(1/2)\,\nu \, {\hat F^{-1}\;  \nu
+i  \nu  \nabla\Lambda}\}\, .\qquad
\label{nu}
\eea
We may then integrate out the $\Lambda$ variables, taking account of the
boundary terms to obtain,
\begin{eqnarray}
U &=&\exp\left\{\int_0^{\,t}\!dt'\!\!\int_{\partial\Omega}\!\!d{\bf s} \cdot
(i\Lambda^a \,  {\bf J})\right\}\! 
\int\!{\cal D}\rho\, {\cal D} { \nu}\delta({\dot \rho}
+\nabla\cdot { \bf J})\, \nonumber \\
&\times & (\det {\hat F})^{-\frac{1}{2}}
\exp\left\{-\frac{1}{2}\int_0^{\,t}\!\!dt'\!\!\int_{\Omega}\!\!d{\bf
  r'}\,  \nu\,  {\hat F}^{-1}\,  \nu\right\} \; , 
\label{langevin}
\end{eqnarray}
where the $\delta$ above is a functional delta function, imposing the Langevin
equation
\begin{subequations}
\label{Langeq}
\begin{eqnarray}
&&{\dot \rho} + \nabla \cdot { \bf J}=0\, , \label{diff0}\\ 
&& { \bf J} =- {\hat D} \nabla\rho + { \nu}\, ,
\label{le}
\end{eqnarray}
with a current noise source ${\nu}$, whose correlator \cite{white} is given by
\be
\langle{ \nu}({\bf r},t) { \nu}({\bf r'},t')\rangle =  
\delta(t-t')\delta({\bf r}-{\bf r}'){\hat F}(\rho) .
\label{noise}
\ee
\end{subequations}
${\bf J}$ may be interpreted as the physical current density 
[not to be confused with the conditional current density (\ref{conslaw1})] 
so that local 
current conservation is guaranteed, and the $(\det {\hat F})^{-1/2}$ serves to
normalize the $\nu$ probability distribution.  The role of the boundary term
is to count the current ${\bf J}$ flowing out of the boundary with the
counting variable $\Lambda^a$, which serves as a Lagrange multiplier.  This
formula gives an immediate translation between the Langevin approach and full
counting statistics, a connection not previously known.  The algorithm is as
follows:
\begin{enumerate}
\item
Given a Langevin equation of the form
(\ref{Langeq}), write the average of the boundary term
with source $\Lambda^a$ as a path integral (\ref{langevin}) over noise and
density fields. \cite{jacobian}

\item
Introduce an auxiliary field $\Lambda$ that takes on the value
$\Lambda^a$ at the boundaries and represents the delta function in
Eq.~(\ref{langevin}) imposing current conservation (\ref{diff0}) in Fourier
form.

\item
Integrate out the Gaussian noise to obtain an action of the form
of Eq.~(\ref{fieldaction}).

\item
Find where the first variation of the action is zero and solve the
equations of motion subject to the boundary conditions.

\item
Insert the solutions back into the action, and do the space and time
integrals.  The answer is the current cumulant generating function.
\end{enumerate}

\section{Perturbation Theory}
\label{cr}

We have shown in Sec.\ \ref{sp} that a large number of participating
elementary charges justifies the saddle point approximation for the generator
of counting statistics.  While the generator may sometimes be found in closed
form,\cite{us} in general, it has no compact expression and the cumulants
should be found separately at every order.  This may be done by expanding
$S_{sp}(Q,\lambda, \chi)$ as a series in $\chi$ and solving the saddle point
equations to a given order in $\chi$ directly.  However, there is another
approach for evaluating the higher cumulants, the cascade diagrammatics
representing higher-order cumulants in terms of the lower ones.  It has been
introduced by Nagaev in the context of mesoscopic charge statistics in the
diffusive wire\cite{Nagaev1} and later extended to the chaotic
cavity,\cite{CavityKirill} but without proof.  The basic idea is that lower
order cumulants mix in to yield corrections to the bare fluctuations of higher
order cumulants.  This method was used successfully in Ref.\
\onlinecite{beenakkerexp} to explain the recent experiment of Ref.\
\onlinecite{reulet}.  In this section, we demonstrate that these rules follow
naturally from the stochastic path integral in the same way as Feynman
diagrams follow from the quantum mechanical functional integral.  In Sec.\
\ref{operator approach} we present another (simpler) method for computing
cumulants based completely on differential operators obtained from the
Hamiltonian equations of motion.  In Sec.\ \ref{netrules} we generalize the
cascade diagrammatics to an arbitrary network, and to the case of
time-dependent correlators.

\subsection{The Principle of Minimal Correlations.}

\begin{figure}[t]
\centerline{\includegraphics[width=7cm]{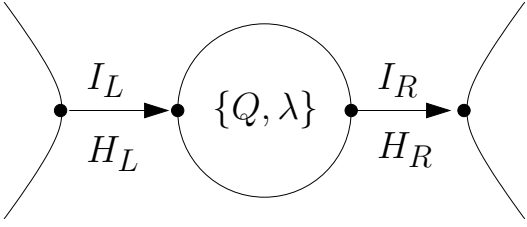}}
\caption{Network representing a chaotic cavity. The state of the internal node
is described by the variable $Q$, the charge on the cavity. The statistics of
the connectors are characterized by the two generating functions $H_{L,R}$.}
\label{MinimalFig}
\end{figure}

To motivate the cascade diagrammatics, we refer to a specific
physical system (see the inset of Fig.\ \ref{Full Statistics}), the
mesoscopic chaotic cavity.\cite{BB} For the purposes of this section,
the cavity is a conserving node carrying charge $Q$, the electronic
reservoirs correspond to the left and right are absorbing nodes, and
the two point contacts are the connectors described by Hamiltonians
$H_L,H_R$ (see Fig.\ \ref{MinimalFig}).  Although a detailed
description of this system is given in Sec.\ \ref{cavityex}, we would
like to mention that the mesoscopic cavity is described by an electron
distribution function $f$, which is fluctuating around its mean value,
$f_0$.  The actual electrical charge in the cavity $Q$ and the
occupation $f$ are related via the large parameter $\gamma$ through
$Q= \gamma(f-f_0)$, where $\gamma = \Delta\mu N_F\gg 1$ (the density
of states at the Fermi energy $N_F$ times the bias $\Delta\mu$) is the
maximum possible number of electrons on the cavity which contribute
to the transport (see Sec.\ \ref{sp}).  

The cascade approach builds on the principle of minimal correlations developed
in Ref.\ \onlinecite{CavityMinCorr}: The point contacts create bare noise
$\langle\langle \tilde{I}_{L}^2 \rangle \rangle = \partial^2 H_L/(\partial i
\lambda_L)^2$, and $\langle\langle \tilde{I}_{R}^2 \rangle \rangle =
\partial^2 H_R/(\partial i \lambda_R)^2$ with no correlation, $\langle\langle
\tilde{ I}_L \tilde{I}_R\rangle\rangle=0$ [see Eq. (\ref{noise3})].  However,
for times longer than the average dwell time of electrons in the cavity, the
current conservation requirement imposes ``minimal correlations'' on the
fluctuations of the physical currents $I_L$ and $I_R$, which can be expressed
in the form of the Langevin equations, 
\be I_{L} = {\tilde I}_{L}- G_L Q, \qquad
I_{R} = {\tilde I}_{R} + G_R Q,
\label{mc}
\ee
where ${\tilde I}_{L,R}$ are now the sources of bare noise, $G_{L,R}$ are the
generalized conductances of the left and right point contact, and $Q$ is the
fluctuating charge in the cavity.  Current conservation of the physical
currents, $I_L= I_R= I$, can now be used to obtain
\be
I = 
\frac{G_R  {\tilde I}_L+G_L  {\tilde I}_R}{G_L+G_R}\;,\qquad
Q = \frac{\tilde{I}_L-\tilde{I}_R}{G_L+G_R} .
\label{mincorr}
\ee
Combining powers of $I$ and $Q$ and averaging over the bare noise, we obtain
the minimal correlation result for arbitrary cumulants $\langle\langle Q^kI^l
\rangle\rangle_m$. In particular, using $\langle\langle{\tilde I}_L {\tilde
I}_R\rangle\rangle=0$, we find the second cumulant of current
is\cite{CavityMinCorr,Langen}
\be
\langle\langle I^2\rangle\rangle =\langle\langle I^2\rangle\rangle_m 
= \frac{G_R^2 \langle \langle
{\tilde I}_L^2 \rangle \rangle+ G_L^2 \langle \langle
{\tilde I}_R^2 \rangle\rangle}{(G_L+G_R)^2}  ,
\label{mcnoise}
\ee
where the subscript $m$ denotes the minimal correlation result.  We stress
that the bare correlators $\langle \langle {\tilde I}_{L,R}^2\rangle\rangle$
are fully determined by the average occupation function $f_0$ of the cavity.

This example demonstrates that a simple redefinition of the current
fluctuations makes it straightforward to find the noise.  Therefore, it came
as a surprise\cite{CavitySchomerus} that the minimal correlation approach is
not sufficient to correctly obtain higher-order cumulants of current.  The
reason for the failure of the minimal correlation approach has been found
recently by Nagaev,\cite{Nagaev1} who showed that from the third order
cumulant on, there are ``cascade corrections'' to the minimal correlation
result, which may be interpreted as ``noise of noise''.  For example, the
third cumulant of current through the mesoscopic cavity,\cite{CavityKirill}
\be
\langle\langle I^3\rangle\rangle = 
\langle\langle I^3\rangle\rangle_m
+ 3 \,\langle \langle I  Q \rangle \rangle_m
\frac{\partial}{\partial Q}\, 
\langle\langle I^2\rangle\rangle_m  ,
\label{mc3}
\ee
contains a contribution from fluctuations of the charge in the cavity that
couples back into the current fluctuations.  The factor of 3 comes from the
fact that there are 3 independent currents that the charge fluctuation may be
correlated with.  For higher cumulants, there will be more cascade corrections
that may be represented in a diagrammatic form.\cite{{Nagaev1,CavityKirill}}

\subsection{Derivation of Diagrammatic Rules.}
\label{rules}

We now present a derivation of these diagrammatic rules for a single node
attached between two absorbing nodes.  Generalizations to an arbitrary network
will subsequently be given in Sec.\ \ref{netrules}.  As we have shown in Sec.\
\ref{sp}, the charge scale imposed by the boundary conditions, $\gamma$, gives
a dimensionless large parameter which justifies the saddle point approximation
of the path integral, so that fluctuations around the saddle point are
suppressed by $1/\gamma$.  In the diagrammatic language, we will show that
loop diagrams are suppressed by the same factor $1/\gamma$.  The diagrammatic
approach given here is based on perturbation theory originally developed
in quantum mechanics. \cite{kleinert}

Consider the path integral expression of the generating function for the
charge absorbed in the left (L) and right (R) node:
\bea
Z(\chi_L,\chi_R)=\int {\cal D}Q{\cal D}\lambda 
\;\exp\Big\{\int_0^{\,t}\!\!dt'[-i {\dot Q}\lambda
\qquad\quad
\nonumber\\
+H(Q,\lambda, \chi_L,\chi_R)]\Big\}  ,
\label{corr-S}
\eea
where $H =H_L(Q, \lambda-\chi_L)+H_R(Q, \chi_R-\lambda)$.  The perturbation
theory is formulated as follows.  First, the external counting variables are
set to zero, $\chi_L=\chi_R=0$.  The Hamiltonian $H \rightarrow
H_L(Q,\lambda)+H_R(Q,-\lambda)$ has a stationary saddle point 
located at $\{Q_0,\lambda_0\}$ that we wish to define as the origin 
of coordinates.  The probability distributions of transferred charge 
are normalized, so
\be
\partial^n_Q H_{L,R}(Q,\lambda)\vert_{\lambda=0}=0, \qquad {\forall} n
.
\label{nor}
\ee 
In particular, $\partial_Q H_L(\lambda)\vert_{\lambda=0}=\partial_Q
H_R(\lambda)\vert_{\lambda=0}=0$, and therefore $\lambda_0=0$.  Next,
$\partial_{i \lambda}H (\lambda)\vert_{\lambda=0}= \langle I_L(Q) \rangle
-\langle I_R(Q) \rangle =0$, since $H_L$ and $H_R$ are the generators of the
left and right current respectively.  Therefore, $Q_0$ is fixed as the charge
in the node such that left and right connector currents are equal on average.
The stability of the saddle point is guaranteed by the fact that the bare
noise correlators, $\langle\langle \tilde{I}_{L,R}^2 \rangle \rangle$, are
positive. The derivatives $\partial_{i \lambda}\partial_Q H_{L}=- G_{L}, \;
\partial_{i \lambda}\partial_Q H_{R}= -G_{R}$ define the generalized
conductance of each connector, where the current flows from left to right in
both connectors.
 
The principle of minimal correlation plays an important role in the
cascade diagrammatics. We will show that this principle is equivalent
to exploiting certain freedoms in the path integral in order to
postpone the cascade corrections to third and higher order cumulants.
In the long-time limit, $t\gg\tau_C$ (where $1/\tau_C=G_L+G_R$ is the
relaxation rate of the charge in the node), the absorbed current is conserved,
$I_R=I_L$. Therefore, the current through the node can be defined as
weighted average of the left and right connector currents $I=(1-v) I_L
+ v I_R$, where $v$ is arbitrary constant. The corresponding counting
variable $\chi$ is introduced by substituting $\chi_R=v\chi$ and
$\chi_L=(v-1)\chi$.  Consider now the second derivative
\be \frac{\partial^2 H}{\partial i\chi \partial Q}{\bigg
\vert}_{\chi=0}= (v-1) G_L + v G_R .
\label{obj}
\ee
We may set it to zero by fixing $v=G_L/(G_L+G_R)$.  This is equivalent to
imposing conservation of current fluctuations as in Eq. (\ref{mincorr}).  If
we consider further the derivative
\be
 \frac{\partial^2 H}{\partial i\lambda \partial Q}{\bigg \vert}_{\chi=0}=
-(G_L + G_R)  ,
\label{obj2}
\ee
we have the freedom to scale $\lambda$ to make the right hand side of
Eq.\ (\ref{obj2}) equal to $-1$ [this scaling only alters the $\chi$
independent prefactor of Eq.~(\ref{corr-S})].  The Hamiltonian takes the new form
\be
H=H_L\left(Q,\frac{G_R \chi +\lambda}{G_L+G_R}\right) 
+ H_R\left(Q,\frac{G_L \chi -\lambda}{G_L+G_R}\right)  .
\label{mccoord}
\ee 
We refer to these new variables as minimal correlation coordinates and
will see that they simplify the diagrammatic expansion.

Define $\delta Q(t)=Q(t)-Q_0$ and $\delta \lambda(t)=\lambda(t)-{\lambda_0}$.
If we expand the Hamiltonian in a power series in $\chi$, $\delta Q$, and
$\delta \lambda$, the terms linear in $\delta Q$ and $\delta \lambda$ vanish
at the saddle point, as well as the $(\delta Q)^2$ coefficient by
Eq.~(\ref{nor}) with $n=2$.  As argued above, in the minimal correlation
coordinates, $\partial_{i\lambda}\partial_Q H(Q_0, \lambda_0)=-1$.  With these
transformations, we may split the action $S$ as
\be
S = S_0+\int_0^{\,t}\!\! dt'V(t'), \quad  
S_0 = -i\int_0^{\,t}\!\!dt'\,\delta \lambda(\tau_C\delta{\dot Q}+\delta Q),
\label{split}
\ee
where $V$ represents the rest of the $H$ power series and will be treated
perturbatively.  It should be emphasized that $V$ is a general nonlinear
function of $\delta \lambda$, so unlike most quantum examples, the full
momentum dependence must be kept.

In order to formulate the perturbation theory, we add two sources, $J$ and $K$
to the action, $S \rightarrow S + \int dt' [J\delta Q +iK \delta \lambda]$, so
that any average of a function of the variables $\delta Q,\delta \lambda$ may
be evaluated by taking functional derivatives with respect to the sources $J$,
and $K$, and then setting the sources to zero.  In particular, for the
generating function we can write
\begin{widetext}
\begin{eqnarray}
Z(\chi) =  \int {\cal D}Q{\cal D}\lambda 
\exp\left\{\int_0^{\,t} dt'\, 
V\left(\delta Q, \delta \lambda, \chi\right)\right\}
\exp\left\{S_0+\int_0^{\,t}dt' [J\delta Q +i K \delta \lambda] 
\right\}{\bigg \vert}_{J,K=0}
\nonumber 
\\
=\exp\left\{\int_0^{\,t}dt'\, V\left(\frac{\delta}{\delta J}, 
\frac{ \delta}{\delta iK}, \chi\right)\right\} 
\int {\cal D}Q{\cal D}\lambda 
\exp\left\{S_0+\int_0^{\,t}dt'[J\delta Q+iK \delta\lambda] 
\right\}{\bigg \vert}_{J,K=0}
\end{eqnarray}
\end{widetext}
Using $S_0$ from Eq.\ (\ref{split}) we evaluate the 
integral over $Q$ and $\lambda$ and obtain:
\be
Z(\chi)=\exp\left\{\int_0^{\,t}\!\! 
dt' V\left(\frac{\delta}{\delta J}, 
\frac{ \delta}{\delta iK},
\chi\right)\right\} W(J,K){\bigg \vert}_{J,K=0}  ,
\label{perout}
\ee
where the functional $W(J,K)$ is
\be
W(J,K) = 
\exp\left\{\int\!\!\int_0^t dt' dt'' J(t') D(t',t'') K(t'')\right\}  .
\label{W}
\ee
The operator $D=(\tau_C\partial_t + 1)^{-1}$ is the retarded propagator, and
may be found explicitly by inverting the kernel in frequency space,
\bea
 D(t,t') &=&
 \int_{-\infty}^\infty \frac{d \omega}{2\pi}
\frac{e^{-i \omega(t-t')}}{-i \tau_C\omega +1}
\nonumber\\
&=&\tau_C^{-1}\Theta(t-t') \exp[-(t-t')/\tau_C]  .
\label{D}
\eea
It describes the relaxation of the charge $Q(t)$ to the stationary state $Q_0$
with the rate $1/\tau_C=G_L+G_R$.

Expanding the exponential in Eq.\ (\ref{perout}) and taking the $t\gg\tau_C$
limit, we arrive at the following expression for the $n$th current cumulant
\bea
\langle \langle I^n\rangle\rangle
=t^{-1}\frac{\delta^n} 
{\delta (i\chi)^n}\sum_{m=1}^{\infty}\frac{1}{m!}  
\left[\int_0^{\,t}\!\! dt'
V\left(\frac{\delta}{\delta J}, 
\frac{\delta}{\delta iK}, \chi \right)\right]^m
\nonumber\\
\times W(J,K){\bigg \vert}_{\stackrel{\chi=J=K=0}{\rm connected}} .
\qquad
\label{sol}
\eea
According to the linked cluster expansion,\cite{Langevin} by considering $\ln
Z(\chi)$ rather than $Z(\chi)$, we have eliminated all disconnected terms.  In
order to compare with the results of Ref.\ \onlinecite{CavityKirill}, we
introduce a new notation by defining
\be
 \partial^j_Q \langle\langle Q^kI^l\rangle\rangle_m \equiv
\partial^j_Q\partial^k_{i\lambda} \partial^l_{i\chi} V(Q_0,\lambda_0, \chi=0)
 .
\label{def}
\ee
Here $\langle\langle Q^kI^l\rangle\rangle_m$ is the irreducible correlator
expressed in terms of the noise sources, i.e. the minimal correlation
cumulant. In this notation, the expansion of $V$ in a Taylor series of all
variables takes the form:
\bea
V(\delta Q,\delta \lambda,\chi)
=\sum_{j,k,l}\frac{1}{j!k!l!}
\partial^j_Q \langle\langle Q^kI^l\rangle\rangle_m
\qquad\qquad
\nonumber\\
\times
[\delta Q(t)]^j[i\delta \lambda(t)]^k[i \chi]^l  .
\label{exphp}
\eea
Inserting the expansion Eq.~(\ref{exphp}) into the formula for the current
cumulants Eq.~(\ref{sol}) gives the formal solution to the problem.  From the
form of $W(J,K)$ and $V$, we can immediately read off the diagrammatic rules
with the internal lines given by the propagators (\ref{D}), and the expansion
coefficients $\partial^j_Q \langle\langle Q^kI^l\rangle\rangle_m$ playing the
role of vertices.

The following simplifications can be done before the rules are finally
formulated. First, it is straightforward to see that loop diagrams are
suppressed by powers of $\gamma^{-1}$. Indeed, according to our
single-parameter scaling assumption, the action (\ref{split}) has a large
prefactor $\gamma$, which can be explicitly displayed, $S\to\gamma\,S$, by
rescaling the charge, $Q\to\gamma\,Q$.  Then it becomes clear that each
propagator $D$, represented by an internal line, comes with a factor of
$\gamma^{-1}$.  Each vertex comes from $V$ and therefore has a factor of
$\gamma$.  If a diagram has $I$ internal lines, $E$ external legs, $V$
vertices and $L$ loops, it will come with a total $\gamma$ power of $V-I$.
Furthermore, Euler's formula tells us that $V+L-I=1$.  Therefore, diagrams
with no loops (``tree'' diagrams) come with a power of $\gamma$, while loop
diagrams are suppressed by the number of loops, $\gamma^{1-L}$. From now on we
will concentrate on tree-level diagrams, since they represent current
cumulants at the level of the saddle-point approximation.

\begin{figure}[t]
\centerline{\includegraphics[width=8cm]{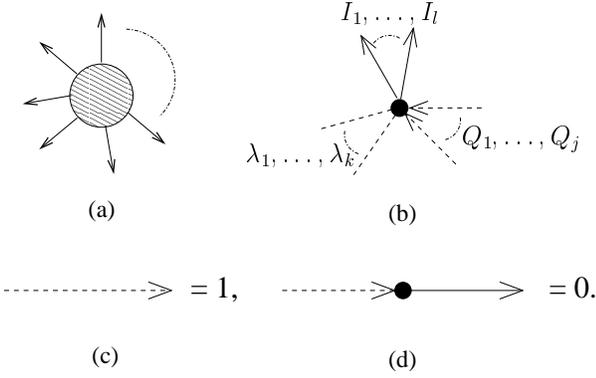}}
 \caption{
{
(a) An $n$-point current cumulant.}
(b) The vertex connecting $l$ external lines with
$j$ internal $Q$ lines and $k$ internal $\lambda$ lines.
(c) The propagator connecting $\lambda$ to $Q$,
equal to $1$ in the stationary limit.
{
(d) The vanishing vertex $\partial_Q\langle I\rangle$
in minimal correlation coordinates.}}
\label{sub-fig-test2}
\end{figure}

Second, in the long time limit, $t\gg\tau_C$, each propagator (\ref{D})
integrated over time gives $1$.  As a result, since every vertex is connected
to at least one other vertex, all the time integrals together simply give a
factor of $t$, and the time dependence cancels on the right hand side of the
Eq.\ (\ref{sol}). There are no time integrals in the vertices and the
propagators just give a factor of $1$ as in Ref. \onlinecite{CavityKirill}. We
are now able to formulate the diagrammatic rules for high-order current
cumulants:
\begin{enumerate}
\item 
The $n$th order cumulant $\langle\langle I^n\rangle\rangle$ is a
connected $n$-point function of $n$ external legs $I$ represented by solid
arrows (see Fig.\ \ref{sub-fig-test2}a).
\item 
The external legs must be connected by using vertices 
(see Fig.\ \ref{sub-fig-test2}b) and linking
internal dashed lines to internal dashed arrows.
\item
The vertices $\partial^j_Q \langle\langle I^lQ^k\rangle\rangle_m$ are
represented by a circle with $l$ external legs, $k$ internal outgoing
dashed lines, and $j$ internal incoming dashed arrows
(see Fig.\ \ref{sub-fig-test2}b).
\item
Multiply each diagram by the number of inequivalent permutations (NIP).
\end{enumerate}
Formally, the vertices $\partial^j_Q \langle\langle
I^lQ^k\rangle\rangle_m$ are the expansion coefficients in
(\ref{exphp}). However, it is important to note that they can also be
easily evaluated by solving the Langevin equations (\ref{mc}) and
expressing the minimal correlation cumulants $\langle\langle
I^lQ^k\rangle\rangle_m$ in terms of cumulants of the noise sources,
$\langle\langle\tilde I_L^{l+k}\rangle\rangle$ and $\langle\langle\tilde
I_R^{l+k}\rangle\rangle$. Some vertices are zero, $\partial^p H/\partial
Q^p(Q_0,\lambda_0)|_{\chi=0}=0$ because of probability conservation,
but other may or may not be zero depending on the physical system.
Here, the advantage of the minimal correlation coordinates is made
clear: the vertex $\partial_Q \langle\langle I\rangle\rangle_m=0$, and
therefore any diagram that contains this vertex is zero (see Fig.\
\ref{sub-fig-test2}d).

To obtain the overall prefactor of a diagram, one can write out all the
numerical constants and count the number of different ways of producing the
same diagram.\cite{Langevin} For example, there is the $n!$ from the $\chi$
derivatives, the $1/m!$ from the Taylor series of $e^V$, a binomial
coefficient from expanding $V^m$, and the $1/(j!k!l!)$ from every vertex with
$j+k+l$ attachments for the different lines.  To compensate these factors, we
have to do the combinatorics of the number of equivalent terms: interchange
the vertices, find the number of different placements of lines on a vertex,
etc.  Often, the number of permutations of the $n$ external legs
will cancel the $m!$, and the $j!k!l!$ number of permutations of the internal
legs attaching to the vertex will cancel that factor arising from the Taylor
expansion.  

Rather than making this expansion, there is a simpler method which
exploits these cancellations given by counting the number of inequivalent
permutations of the diagram (NIP).  
The NIP of the diagram
is defined by how many ways the external legs of the diagram
may be relabeled, such that the diagram is not topologically equivalent under
deformation of the external legs.  In other words,
a diagram with $n$ external
legs has $n!$ ways of labeling them.  If this diagram with a given
labeling of the legs may be topologically deformed to
give the diagram back with a different labeling, these two sets
of labelings are equivalent permutations. 
If we write out all the different labelings the external legs
can have, and cross out every labeling that is an equivalent
permutation of another, then the number of labelings that remain 
is the NIP.
This number is most easily found by dividing $n!$
by the number of equivalent permutations of the diagram.
The number of equivalent permutations of the diagram is also called
the symmetry factor of the diagram.

\begin{figure}[t]
\centerline{\includegraphics[width=8cm]{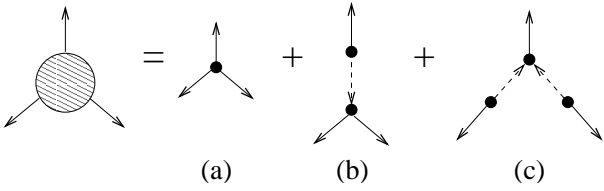}}
\caption{Tree level contributions to the third cumulant of
transmitted current.}
\label{cumulant3}
\end{figure}

We illustrate these two approaches with the third cumulant, see Fig.\
\ref{cumulant3}.
With the simplifications discussed above, these diagrams may be written as 
\bea
\langle\langle I^3 \rangle\rangle =  
\langle\langle I^3 \rangle\rangle_m  
&+&3 \langle\langle I\, Q  \rangle\rangle_m
 \frac{\partial}{\partial Q} \langle \langle I^2 \rangle\rangle_m 
 \nonumber\\
&+&3\langle \langle I\, Q \rangle\rangle_m^2\, \frac{\partial^2}{\partial
  Q^2}\langle\langle I \rangle \rangle_m\; .
\label{trdcum}
\eea
Note that diagram (c) does not appear in Ref.\
\onlinecite{CavityKirill}, because it happens to vanish for the
chaotic cavity [see also Eq.~(\ref{mc3})].  Referring to the formula
(\ref{sol}), the contributions in Eq.~(\ref{trdcum}) are from
$m=1,2,3$ respectively.  Each diagram must have a $\chi^3$ term in the
expansion.  We first show the combinatorial method to obtain the
prefactor: Diagram (a) has a factor of 1/3! from the number of
permutations of the $\chi$ variables, canceling the 3! from the $\chi$
derivatives.  Diagram (b) has a factor of 1/2! from the number of
permutations of the $\chi$ variables, a factor of 1/2! from the Taylor
series of the exponential, a factor of 2 from the binomial expansion
of $V^2$, and the 3! from the $\chi$ derivatives, leaving a factor of
3.  Diagram (c) has a factor of 1/3! from the Taylor series of the
exponential, a factor of 3 from the binomial expansion of $V^3$, a
factor of 1/2! from the number of permutations of the $\delta Q$
variables, a factor of 2 from the functional derivatives acting on
$W$, and the 3! from the $\chi$ derivatives, leaving a factor of 3.
The NIP is simpler to derive: We divide the number of
permutations of the external legs, $m!$, by the number of equivalent
permutation of the elements of the diagram that leave it unchanged.  The
number of equivalent permutations of diagrams (a,b,c) are 3!, 2!, 2!, leaving
the overall factors 1, 3, 3.
\begin{figure}[b]
\centerline{\includegraphics[width=8cm]{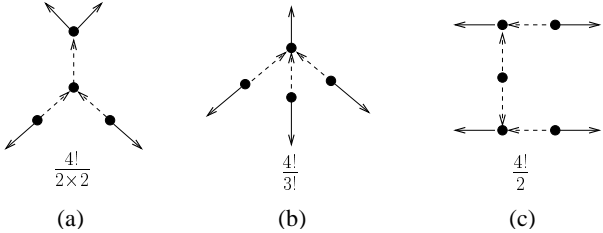}}
 \caption{Three examples of diagrams contributing to the fourth 
cumulant.}
\label{dia4}
\end{figure}

The computation of these diagrammatic contributions is best understood
by a little practice on some examples.  Consider three of the diagrams that
contribute to the fourth cumulant drawn in Fig. \ref{dia4}.
The diagrams symbolically represents the combinations:
\begin{subequations}
\begin{eqnarray}
(a) &=&  
\frac{\partial}{\partial Q}  \langle \langle I^2 \rangle\rangle_m 
 \frac{\partial^2}{\partial Q^2}\langle\langle Q \rangle \rangle_m
   \langle \langle I\, Q \rangle\rangle_m^2\;,
\label{fourthd1} \\
(b) &=&  
 \frac{\partial^3}{\partial Q^3}\langle\langle I \rangle \rangle_m
   \langle \langle I\, Q \rangle\rangle_m^3\;,
\label{fourthd2}\\
(c) &=& \langle\langle Q^2 \rangle \rangle_m
 \left(\frac{\partial^2}{\partial Q^2}\langle\langle I \rangle
 \rangle_m\right)^2 \langle \langle I\, Q \rangle\rangle_m^2\;.
\label{fourthd3}
\end{eqnarray}
\end{subequations}
To figure out the numerical prefactors, we divide $4!$ (4 is the number of 
external legs) by the symmetry factor of the diagram. We first consider 
the symmetry factor of (a): The upper two legs may be flipped, and the lower two
legs may be independently flipped where the dotted arrows join without
altering the topology of the diagram.  Therefore, the symmetry factor
is $2\times2=4$, and the NIP is $4!/4  = 6$. Moving on to diagram (b), 
the three lower legs may be permuted amongst themselves to give a symmetry 
factor $3!$, and therefore the NIP is $4!/3!=4$.  
Finally, diagram (c) may be flipped about its center for a symmetry factor 
of $2$, giving a NIP of $4!/2=12$.

\subsection{Operator approach.}
\label{operator approach}

In the stationary limit, $t\gg \tau_C$, the action takes the form $S=
t H(Q, \lambda, \chi)$ so that the evaluation of the cumulant
generating function reduces to finding the stationary point of the
Hamiltonian $H$ as a function of variables $\lambda$ and $Q$.  This
can be done by solving equations ${\partial_Q}H=0$ and
$\partial_{\lambda}H=0$. The generating function is then obtained by
substituting the solutions $\{{\bar Q}$, ${\bar \lambda}\}$ into the
Hamiltonian. In the previous section we have shown that this problem
can be solved using path integral methods, and the solution can be
represented diagrammatically. In the next section we will exploit the
full strength of the path integral formalism in order to generalize
the diagrammatics to an arbitrary network, and for the case of
time-dependent charges. However, in the stationary limit, the
conceptual simplicity of the problem of finding the stationary point
of the function $H$ indicates that there should exist a simple
iterative procedure for evaluating the cumulants up to a given
order. In this section we use classical mechanics methods to prove
that this is indeed the case.

We first make the variable transformation $i\lambda\rightarrow \lambda$, and
$i\chi \rightarrow \chi$, so that the Hamiltonian becomes a real function. For
$\chi=0$ the saddle point is located at $\{Q_0,\lambda_0\}$.  For non-zero
$\chi$ the saddle point moves to a new position $\{\bar Q,\bar\lambda\}$,
which depends on $\chi$, and the Hamiltonian $H(\bar Q,\bar\lambda, \chi)$
becomes the generator of cumulants of the current,
\be
\langle\langle I^n\rangle\rangle=d^nH(\bar Q,\bar\lambda,\chi)/d\chi^n|_{\chi=0}.
\ee
By expressing the total $\chi$ derivative in terms of partial
derivatives, the average current can be written as 
\be
\langle I\rangle=
(\partial_\chi+Q' \partial_{Q} +\lambda' \partial_{\lambda})
H(Q,\lambda,\chi)|_{\{\chi=0,Q_0,\lambda_0\}},
\label{average}
\ee
where $Q'=dQ/d\chi, \lambda'=d\lambda/d\chi$ are $\chi$ dependent.  We
wish to eliminate the functions $Q'$ and $\lambda'$ and to express the
cumulant in terms of the partial derivatives of $H$.  This is done by
applying a total derivative to the equations of motion: $[\partial_Q
H]'=[\partial_\lambda H]'=0$ and leads to two equations for $Q'$ and
$\lambda'$ which may be solved,
\be
Q'=\frac{\{
\partial_\lambda H, \partial_\chi H
\}}{\{
\partial_Q H,\partial_\lambda H
\}},\quad
\lambda'=-\frac{\{
\partial_Q H, \partial_\chi H
\}}{\{
\partial_Q H,\partial_\lambda H
\}},
\label{brackets}
\ee
where $\{A,B\}$ is the Poisson bracket, defined as $ \{A,B\}= \partial_\lambda
A\, \partial_Q B -\partial_Q A\, \partial_\lambda B$.  The solutions have to
be inserted into the Eq.\ (\ref{average}).

The advantage of this representation is clear: Now the right hand side of the
Eq.\ (\ref{average}) (before taking the $\chi=0$ saddle point) depends only on
variables $\lambda$, $Q$, and $\chi$. Therefore, we can apply the
procedure again in order to express the high-order cumulant in terms of
partial derivatives.  This procedure solves the problem by giving a single
operator,
\be
{\mathrm D} = \partial_\chi+\frac{\{\partial_\lambda H, \partial_\chi H\}
 \partial_Q - 
\{\partial_Q H, \partial_\chi H\} \partial_\lambda}{\{\partial_Q H,
  \partial_\lambda H\} },
\label{diff}
\ee
which, being applied $n$ times to a given Hamiltonian $H$ and evaluating the
resulting expression at the $\chi=0$ saddle point, gives cumulants of current:
\be
\langle \langle I^n \rangle \rangle = {\mathrm D}^n H(Q,\lambda, \chi)|_
{\{\chi=0,Q_0,\lambda_0\}}. 
\label{cum}
\ee
This approach is obviously more simple compared to the diagrammatic method,
since in the diagrammatics, after drawing all of the diagrams, they have to be
evaluated individually by taking many partial derivatives of the Hamiltonian
and evaluating them at the $\chi=0$ saddle point.  With this new approach,
given the Hamiltonian $H$, the operator ${\mathrm D}$ may be constructed
(\ref{diff}) and with a mathematical program, an arbitrary cumulant may be
easily computed (\ref{cum}).

It is easy to see the importance of the minimal correlation coordinates in
this solution.  After applying ${\mathrm D}$ several times, the derivative
quotient rule generates a large number of denominators, $\{\partial_Q H,
\partial_\lambda H\}= (\partial_Q \partial_\lambda H)(\partial_\lambda
\partial_Q H) -(\partial_Q\partial_Q H)(\partial_\lambda \partial_\lambda H)$.
At $\chi=0$, as we argued previously, $\partial_Q\partial_QH=0$, and it is
possible to change coordinates so that $\partial_Q \partial_\lambda H=-1$.  As
a result, the denominator in (\ref{cum}) is equal to $1$, which greatly
simplifies the expansion. Finally, we would like to stress that the operator
approach, introduced in this section for the one node case, can be easily
generalized to a network.

\subsection{Network Cascade Diagrammatics: 
Correlation Functions.}
\label{netrules}

Consider now a general network.  In the Sec.\ \ref{rules}, we saw that the
dominant contribution to Eq.~(\ref{corr-S}) arises from tree-level diagrams.
On time scales $t\gg \tau_C$, the time dependence drops out, and the current
cumulants are static.  We now generalize the diagrammatic rules presented in
the section \ref{rules} to investigate time- and node-dependent correlation
functions of conserved and absorbed charges, Eq.\ (\ref{corr}).  To define the
network, we must arbitrarily label the current flow, yielding a directed
network.  By doing so we fix the signs of the elements
$H_{\alpha\beta}=-H_{\beta\alpha}$ of the Hamiltonian. In particular, the
elements of the generalized conductance matrix $\hat G$,
\be
G_{\alpha\beta}=
\frac{\partial^2H}{\partial(i\lambda_\alpha)\partial Q_{\beta}}\; ,
\ee
(evaluated at ${\bf Q}={\bf Q}_0,{\bf \Lambda}=0$) are negative or positive
depending on the chosen direction.  If we segregate absorbing (a) and
conserving (c) nodes, the conductance matrix $\hat G$ may be put in block
form.  Two of them, the blocks $\hat G_{cc}$ (real symmetric) 
and $\hat G_{ac}$ will be
relevant. This gives us the necessary tool to define the generalized minimal
correlation coordinates.  We consider the frequency dependent response by
letting the evolution time extent to infinity, and introduce the time Fourier
transform of the variables $\{ {\bf Q}^c,{\bf \Lambda}^c,{\bf \chi}^c, {\bf
\chi}^a\}$, where the vector $\{{\bf \chi}^c,{\bf \chi}^a\}$ is a
time-dependent source term introduced to produce correlation functions of the
conserved and absorbed currents [see Eq.\ (\ref{corr})].

Following the steps of section \ref{rules}, we again split the action into two
parts, $S=S_0+\int dt\,V$, where
\begin{eqnarray}
S_0&=&i \int dt [- {\bf \Lambda}^c {\bf \dot Q}^c
+ {\bf \Lambda}^c\hat G_{cc}{\bf  Q}^c  + 
({\bf \chi}^c\hat G_{cc}+{\bf \chi}^a\hat G_{ac}){\bf  Q}^c]
\nonumber  \\
&=&i\int\!\!\int\frac{d\omega d\omega'}{2\pi }\;
[ {\bf \Lambda}^c(i\omega'+\hat G_{cc}){\bf  Q}^c 
\nonumber  \\
&{}&\qquad\qquad\quad 
+ ({\bf \chi}^c\hat G_{cc}+
{\bf \chi}^a\hat G_{ac}){\bf  Q}^c]\,
\delta(\omega+\omega'), 
\label{s0net}
\end{eqnarray}
and where we have dropped the $\delta$ in front of the variables 
for simplicity.  As in
Sec.\ \ref{rules}, the generalized minimal correlation coordinates are defined
by shifting and rescaling the ${\bf \Lambda}^{c}$ variables in order to
eliminate the ${\bf \chi}$ variables in Eq.\ (\ref{s0net}).  However, because
${\bf \chi}$ is now a vector, the proportionality factor must be a frequency
dependent matrix,
\be
{\bf \Lambda}^{c}(\omega) \rightarrow 
\hat D^\dagger(\omega)[{\bf \Lambda}^{c}(\omega)
+ \hat G^\dagger_{cc}\chi^c(\omega)+
\hat G^\dagger_{ca}\chi^a(\omega)].
\label{shiftnet}
\ee
Here $\hat D(\omega)$ is the matrix network propagator,
\be
\hat D(\omega)=-(i\omega\hat E+\hat G_{cc})^{-1},
\label{networkprop}
\ee
and $\hat E$ is the identity matrix.  It is straightforward to verify that
after the shift, the functional $\int dt\,V$ becomes the generator of
cumulants of minimal correlation currents, i.e.\ of the currents which are
solutions of the Langevin equations:
\begin{subequations}
\label{LE}
\bea
I_\alpha^c&=&-i\omega Q^c_{\alpha}=
-i\omega \sum_{\beta\gamma} D_{\alpha\beta}(\omega){\tilde I}_{\beta\gamma}
\\
I_\alpha^a&=&\sum_{\beta\gamma\alpha'}G_{\alpha\alpha'}
D_{\alpha'\beta}(\omega){\tilde I}_{\beta\gamma}
+\sum_{\gamma}{\tilde I}_{\alpha\gamma}\,,
\eea 
\end{subequations}
where ${\tilde I}_{\alpha\beta}$ are the bare noise sources as defined in
Eq. (\ref{noise3}).  We finally rescale
$\chi^c(\omega)\to\chi^c(\omega)/(i\omega)$ in order to replace conserved
currents with charges, ${\bf I}^c\to {\bf Q}^c$.

The total action now acquires the following form
\bea
S &=& (2\pi i)^{-1}\int d\omega\,
{\bf \Lambda}^c(-\omega){\bf  Q}^c(\omega)
\nonumber\\
&+&\int\!\! dt V\!\left[{\bf Q}^c,
\hat D^\dagger({\bf \Lambda}^{c}+ \chi^c)+
\hat D^\dagger\hat G^\dagger_{ca}\chi^a,
{\bf \chi}^a\right],\;
\label{stotnet}
\eea 
where the simplified form of the ${\bf \Lambda}$ argument of $V$ follows after
composing the various transformations.  Following the plan of the previous
section, we replace the charge and counting variables $\{{\bf Q}(\omega), {\bf
\Lambda}(\omega)\}$ by functional derivatives with respect to the charge and
counting sources $\{{\bf J}(\omega),{\bf K}(\omega)\}$, and take the $V$ term
outside of the functional integral.  The functional integrals may now be
performed to obtain
\be W({\bf J},{\bf K}) = \exp\left\{\int\!\!\int\frac{d\omega d\omega'}
{2\pi}\,{\bf J}(\omega'){\bf K}(\omega)\delta(\omega+\omega')\right\}.
\label{W2}
\ee
The perturbation $V$ must now be expanded in a Taylor series with respect to
all variables.  The time dependence only appears through the variables
themselves, so the expansion coefficients will be time independent, with the
exception of the propagator $D_{\alpha\beta}(\omega)$ multiplying the counting
variables.
\begin{widetext}
\bea
V =&& \sum_{\{i_\alpha, j_\alpha, k_\alpha,l_\alpha\}=0}^\infty 
\frac{\delta^{j_1+\cdots+j_n}}{\delta (Q_1^c)^{j_1}\cdots
\delta (Q_n^c)^{j_n}}
\langle\langle 
(I_1^a)^{l_1} \cdots (I_r^a)^{l_r}
(Q_1^c)^{i_1} \cdots (Q_q^c)^{i_q}
(Q_1^c)^{k_1} \cdots (Q_p^c)^{k_p}
\rangle\rangle_m\nonumber\\
&&\times \frac{(\chi_1^a)^{l_1}}{l_1!}\cdots\frac{(\chi_r^a)^{l_r}}{l_r!}
\times \frac{(\chi_1^c)^{i_1}}{i_1!}\cdots\frac{(\chi_r^c)^{i_q}}{i_q!}
\times\frac{ \lambda_1^{k_1}}{k_1!}\cdots\frac{ \lambda_p^{k_p}}{k_p!}
\times\frac{ (Q_1^c)^{j_1}}{j_1!}\cdots\frac{ (Q_n^c)^{j_n}}{j_n!}\; .
\label{bigV}
\eea
\end{widetext}
As in the one node case, the vertices  $\delta_{Q^c_{\alpha}}
\langle I^a_{\beta}\rangle$ vanish.
We note again that the notation chosen for the expansion coefficients in
Eq.~(\ref{bigV}) connects the formalism described here with the Langevin
equation point of view.  The minimal correlation cumulant
$\langle\langle\ldots \rangle\rangle_m$ may be calculated 
either by the expansion procedure described by
Eqs.~(\ref{stotnet},\ref{bigV}), or by expressing
the physical currents and charges in terms of the current source cumulants by
solving the Langevin equations for currents and charges, given by Eq.\
(\ref{LE}).

The $n$th order irreducible correlator $\langle\langle I^a_1(\omega_1) \cdots
Q^c_n(\omega_n)\rangle\rangle$ may be expressed as a tree-level diagram with
$n$ external lines representing absorbed currents $I^a_\alpha$ and conserved
charges $Q^c_\alpha$.  Every vertex is local in time, so if there are $p$ legs
at a vertex, each is assigned an independent frequency, while the time
integral imposes overall frequency conservation, $\delta(\sum_i \omega_i)$.
The cascade rules are generalized as follows:
\begin{enumerate}
\item
Every vertex represents the object
$$\delta_{Q_1^c(\omega_1)}\cdots\delta_{Q_l^c(\omega_l)} \langle\langle
I^a_{l+1}(\omega_{l+1})\cdots Q^c_n(\omega_n) \rangle\rangle_m\,,$$ which is
multiplied by a $\delta$-function conserving overall frequency,
$\delta(\sum^n_{i=1}\omega_i)$.
\item 
The minimal correlation cumulants $\langle\langle
I^a_{l+1}(\omega_{l+1})\cdots Q^c_n(\omega_n) \rangle\rangle_m$
may be evaluated by expressing them in terms of cumulants of sources 
$\langle\langle\tilde I_{\alpha\beta}^n\rangle\rangle$ via the solutions
(\ref{LE}) of the Langevin equations, or by Eq.(\ref{bigV}) if the
Hamiltonian is known.
\item
The internal dashed arrow goes from $Q^c_\alpha(\omega)$ to
$\delta_{Q^c_\alpha(\omega)}$. It conserves the node index $\alpha$ and the
frequency $\omega$.\cite{commentrule}
\item
External lines for absorbed currents and conserved charges originate from
$I^a_\alpha(\omega)$ or $Q^c_\alpha(\omega)$ of the vertexes.  They conserve
the node index and the frequency.
\item
Sum over all internal node indices, and integrate over all internal
frequencies to remove all but one of the frequency delta functions.
\item
The result has to be multiplied by the total number of inequivalent
permutations.
\end{enumerate}

The cascade rules are easily extended to the field theory (see Sec.\
\ref{ft}).  The functional analog to the inverse conductance matrix is
the operator
\be 
\hat G^{-1}({\bf r}-{\bf r}')  \equiv \frac{\delta^2 h}{\delta\lambda({\bf r})
\, \delta\rho({\bf r}')}=
-\delta({\bf r}-{\bf r}')\nabla\hat D \nabla\; .
\label{prop}
\ee
The diffusion propagator $(i\omega +\hat G^{-1})^{-1}$ can be used to solve
the Langevin equations (\ref{Langeq}) for the density
$\rho(\omega, {\bf r})$ and current $I(\omega)$ in order to evaluate minimal
correlation cumulants. We would like to stress that these cumulants are
limited to second order only, because in the diffusion limit the noise sources
are Gaussian. The summation over node indices is replaced with an integration
over the coordinate ${\bf r}$.

\section{Applications}
\label{app}

The formalism presented above is intentionally abstract and general. This is
to facilitate maximum applicability and not to tie it to a particular field.
However, it is important to give concrete examples.  For this reason, we give
a detailed treatment of two problems.  As a first problem, we consider the
saddle-point equations of the 1D field theories for $D$ and $F$ being
arbitrary functions of the density $\rho$ [see Eq.~(\ref{act1d})]. We apply
the results of this analysis to the transport in a diffusive mesoscopic wire
at zero temperature, rederive the FCS generating function of the transmitted
charge obtained in Refs.\ \onlinecite{wire} and \onlinecite{Nazarov1}, and
give new results. We also prove the conjecture made in Ref.\
\onlinecite{derrida} that the current noise of the diffusive symmetric
exclusion process at half-filling is Gaussian, i.e.\ all high-order cumulants
of transmitted charge vanish.  In the end of the Sec.\ \ref{wire} we
generalize our results to multi-dimensional diffusion models and prove the
universality of their transport statistics.  As a second problem, we address
the statistics of charge fluctuations in a mesoscopic chaotic cavity. We
explicitly find the probability distributions for different physical
configurations.

\subsection{FCS for one-dimensional field theories. The mesoscopic 
diffusive wire.}
\label{wire}

Before demonstrating our solution for the FCS of the mesoscopic diffusive wire
specifically, we first consider the general 1D field theory with the action
(\ref{act1d}). In the stationary limit, ${\dot \rho} = {\dot \lambda}=0$ the
action can be written as
\be
S=t\int\limits_0^Ldz \left[-D\rho'\lambda'+\frac{1}{2}F(\lambda')^2\right].
\label{action-stat1}
\ee
The stationary saddle-point equations
\begin{equation}
(F\lambda'-D\rho')'=0, \quad
2D\lambda''+\frac{\delta F}{\delta\rho}\,(\lambda')^2=0,
\label{spe12}
\end{equation}
can be partially integrated leading to the 
following two equations:
\begin{subequations}
\label{parsol}
\be
D\rho'=\pm\sqrt{{\cal I}^2-2{\cal H}F},
\label{sol1}
\ee
\be
\lambda'=2{\cal H}/({\cal I}-D\rho').
\label{sol2}
\ee
\end{subequations}
The two integration constants ${\cal I}=-D\rho'+F\lambda'$ and ${\cal
H}=-D\rho'\lambda'+(F/2)(\lambda')^2$ are the 
conserved (conditional) current and
the Hamiltonian density, respectively.  These conservation laws follow from
the symmetries of our $1D$ field theory
[see Eqs.\ (\ref{conslaw1}) and (\ref{conslaw2}) and the surrounding discussion].  Thus we
obtain the following result for the action (\ref{action-stat1}),
\be
S=tL{\cal H}.
\label{action-stat2}
\ee
The equations (\ref{parsol}) and (\ref{action-stat2}) represent
the formal solution of the FCS problem for 1D diffusion models with $D(\rho)$
and $F(\rho)$ being arbitrary functions of $\rho$.  The following procedure
has to be done in order to obtain the cumulant generating function $S(\chi)$
of the transmitted charge:
\begin{enumerate}
\item 
The differential equation (\ref{sol1}) has to be solved for $\rho(z)$ with the
boundary conditions $\rho(z)|_{z=0}=\rho_L$ and $\rho(z)|_{z=L}=\rho_R$. The
constant ${\cal I}$ should be expressed through the constants $\rho_L$,
$\rho_R$, and ${\cal H}$.
\item 
Next, $\rho(z)$ is substituted into Eq.~(\ref{sol2}) which is integrated to
obtain $\lambda(z)$ with the boundary conditions $\lambda_L=0$ and
$\lambda_R=\chi$.
\item 
Finally, using the solution for $\lambda(z)$ the constant ${\cal H}$ is
expressed in terms of $\rho_L$, $\rho_R$, $\chi$, and substituted into the
action (\ref{action-stat2}).
\end{enumerate}
We note that by expressing ${\cal H}$ and $\chi$ in terms of ${\cal I}$, 
we may also formally obtain the logarithm of the current distribution, 
\be
\ln P(I)=S({\cal I})-t{\cal I}\chi({\cal I}),\quad {\cal I}\to I,
\label{LNPI}
\ee
as a result of the stationary phase approximation for the integral $P(I)=\int
d\chi\exp[S(\chi)-tI\chi]$ and because $\partial {\cal H}/\partial\chi={\cal
I}/L$.

As an example of the 1D field theory, we consider the FCS of the
electron charge transmitted through the mesoscopic diffusive
wire. When the chemical potential difference $\Delta\mu=\mu_L-\mu_R>0$
is applied to the wire, the electrons flow from the left lead to the
right lead with the average current $I_0=e^{-1}G\Delta\mu$, where $G$
is the conductance of the wire. The elastic electron scattering causes
non-equilibrium fluctuations of the current.  At zero temperature, and
for noninteracting electrons (the cold electron regime), the
zero-frequency current noise power has been
found\cite{onethird,kirillrate,Henny} to be equal to $\langle\langle
I^2\rangle\rangle=(1/3)eI_0$, i.e.\ the noise is suppressed compared
to the Poissonian value. The suppression factor $1/3$ was shown to be
universal,\cite{Nazarov2,Sukhorukov} i.e.\ it does not depend on the
character of the disorder or on the shape of the wire. The FCS of the
transmitted charge has been studied in Refs.\ \onlinecite{wire} and
\onlinecite{Nazarov1} using quantum-mechanical methods, and recently
in Ref.\ \onlinecite{derrida} using a classical method with the
following result for the generating function of cumulants of the
dimensionless charge $Q/e$:
\be
S(\chi) = (tI_0/e)\,{\rm arcsinh}^2 \left[\sqrt{\exp(\chi)-1}\,\right].
\label{ans}
\ee
Here we will rederive this result using our classical method.
  
On the classical level, the electrons in the diffusive wire are
described by the distribution function $f(z)$. 
Under 
transport conditions (and at zero temperature), this distribution $f(z)$
varies from $f_L=1$ in the left lead to
$f_R=0$ in the right lead. Starting from the Langevin equation\cite{Langevin}
as described in Sec.\ \ref{ft} or, alternatively, taking the continuum limit
for the series of mesoscopic cavities,\cite{Ob2} we arrive at the action
(\ref{action-stat1}) with the form\cite{commentScr}
\be S=(tI_0/e)\int\limits_{-1/2}^{1/2}dz
[-f'\lambda'+f(1-f)(\lambda')^2],
\label{action-stat3}
\ee
where we have rescaled the coordinate $z$, $\rho(z)$ has been replaced with
the distribution $f(z)$, and where $D=1$, and $F=2f(1-f)$ up to the overall
constant $I_0/e$.  This form of $F$ is quite general for fermionic systems. It
originates from the Pauli blocking factors, i.e. the transition probability is
proportional to the probability that the initial state is populated times to
probability that the final state is empty.\cite{kirillrate} Applying now the
procedure described in the beginning of this section, we solve the
saddle-point equations and find the fields $f$ and $\lambda$,
\begin{subequations}
\label{soldiff}
\be
f(z,\chi)=\frac{1}{2} \left[ 1- \frac{\sinh (2\alpha z)}
{\sinh\alpha} \right] ,
\label{f} 
\ee
\be
\lambda (z,\chi) = 2 \, {\rm arctanh} \left[\tanh (\alpha/2) 
\tanh(\alpha z) \right] ,
\label{l} 
\ee
\be 
\alpha = {\rm arcsinh}\left[\sqrt{\exp(\chi)-1}\,\right],
\label{c2} 
\ee
\end{subequations}
where ${\cal H}=\alpha^2$, so that according to the Eq.\ (\ref{action-stat2})
we immediately obtain the result (\ref{ans}). 

\begin{figure}[t]
\centerline{\includegraphics[width=7cm]{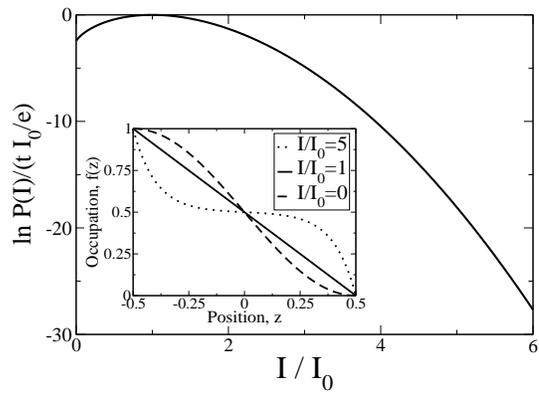}}
\caption{The logarithm of the distribution of the current
  through a mesoscopic diffusive wire as a function of the ratio 
  $I/I_0$ of the current to its average value $I_0$. The distribution
  is strongly asymmetric, with the Gaussian tale at $I\gg I_0$. Inset:
  The electron occupation $f$ inside the wire as a function of the 
  rescaled coordinate $z$, under the condition that the average current 
  $I=I_0$, no current $I=0$, and large current $I=5I_0$ has been measured.}
  \label{sub-fig-test}
\end{figure}

The logarithm of the current distribution $\ln[P(I)]$ can be now found from
the equation (\ref{LNPI}). We obtain the following result: 
\be
\ln[P(I)]=-(tI_0/e)[2\alpha\coth\alpha\ln(\cosh\alpha)-\alpha^2], \ee where
$\alpha$ has to be expressed in terms of ${\cal I}=I/I_0$ by solving the
equation \be \alpha\coth\alpha=I/I_0.
\label{alpha}
\ee
The last equation has real positive solutions, $0<\alpha<\infty$, for $I>I_0$,
and pure imaginary solutions $\alpha=i\beta$ with $0<\beta<\pi/2$, for
$I<I_0$.  The distribution $P(I)$ is strongly asymmetric around the average
current $I=I_0$ (see Fig.\ \ref{sub-fig-test}). It has the following
asymptotics: $\ln P=-(tI_0/e)[{\cal I}^2-(2\ln2){\cal I}]$, for ${\cal
I}=I/I_0\gg 1$, i.e.\ $P$ has a Gaussian tail, and $\ln P=-(\pi^2/4)(tI_0/e)$,
for $I=0$.

We also plot the conditional electron occupation $f(z, I)$, Eq.~(\ref{f}), for
different values of the normalized current $I/I_0$. There are several
interesting points to stress.  (i) For large currents, $I>I_0$, the
function $f$ drops mostly at the ends of the wire, while for small currents,
$I<I_0$, the drop of $f$ is mostly concentrated in the center of the
wire. This effect has a simple explanation. At the end points of the wire,
$z=\pm1/2$, the occupation $f(z)$ is fixed independent of the particular value
of the current $I$. On the other hand, its derivative takes the value
$f'=-{\cal I}=-I/I_0$ at $z=\pm1/2$, which can be easily verified using Eqs.\
(\ref{f}) and (\ref{alpha}).  As a result, $f(z)$ deviates from its linear
behavior, $f(z)=1/2-z$, characteristic of the average value of current,
${\cal I}=1$.  The actual reason for this effect is that according to Eq.\
(\ref{le}) the total current ${\cal I}=-f'+\nu$ contains a contribution from
the source of noise, $\nu$. The greatest contribution is concentrated at the
center of the wire, where the noise power $F=2f(1-f)$ has its maximum, while
it vanishes at the ends of the wire. Since the current ${\cal I}$ is
conserved, $f'$ has to be redistributed in such a way as to partially
compensate the effect of the source $\nu$.  (ii) Fluctuations of $f$ are
strongly suppressed at the ends of the wire, which is imposed by the boundary
conditions, and at the center of the wire, as a result of the discrete
symmetry, $\{z\to-z;f\to1-f\}$.  (iii) Eq.\ (\ref{alpha}) has
additional solutions with $\beta>\pi/2$.  These solutions are not physical
however, since $f$ becomes negative or larger than $1$ leading to $I<0$, which
is impossible at $T=0$.

Returning to the saddle-point equations (\ref{spe12}), we note that if $\delta
F/\delta\rho=0$ for a particular density $\rho_0$, then the fields
$\rho(z)=\rho_0$, and $\lambda(z)=\chi z/L$ solve these equations. The
fluctuations of the current become Gaussian with the noise power
$\langle\langle I^2\rangle\rangle=F(\rho_0)/L$. This generalizes and proves 
the conjecture
made in Ref.\ \onlinecite{derrida} that the noise of the diffusive symmetric
exclusion process is Gaussian at half-filling, $f=1/2$.

As a final remark we note that the whole class of multi-dimensional field
theories,
\be
S = t\int_\Omega d{\bf r} \,[-\nabla{\lambda} {\hat D}\nabla{\rho} +
(1/2)\nabla{\lambda}{\hat F}\nabla{\lambda}],
\label{action-stat4}
\ee
with ${\hat D}=D(\rho)\,{\hat T}$, ${\hat F}=F(\rho)\,{\hat T}$, and $\hat T$
being an arbitrary constant symmetric tensor,\cite{commentT} 
bear the same kind of the
universality as the shot noise in diffusive conductors discussed above (see
Refs.\ \onlinecite{Nazarov2} and \onlinecite{Sukhorukov}).  The reason is that
the field theory with the action (\ref{action-stat4}) can be mapped on the 1D
theory with the action (\ref{action-stat1}) by making use of the
parameterization
\begin{equation}
\rho({\bf r})= \rho\,[\varphi({\bf r})],\quad 
\lambda({\bf r})= \lambda[\varphi({\bf r})], 
\label{fields}
\end{equation}
where the function $\varphi({\bf r})$ satisfies the equation
\be
\nabla\cdot[\hat T\nabla \varphi({\bf r})]=0.
\label{varphi}
\ee
Using Eqs.\ (\ref{spe12}) for $\rho$ and $\lambda$ as functions of $\varphi$,
it is straightforward to verify that the fields $\rho({\bf r})$ and
$\lambda({\bf r})$ given by (\ref{fields}) and (\ref{varphi}) satisfy the
saddle-point equations for the action (\ref{action-stat4}).  One of the
equations is the conservation of current:
\be
{\bf j}=-\hat D\nabla\rho+\hat F\nabla\lambda={\cal I}\,\hat T\nabla\varphi.
\label{current-density}
\ee
Since the 1D Hamiltonian density is conserved, the action takes the following
form
\begin{equation}
S=t{\cal G}{\cal H},
\label{action-stat5}
\end{equation}
where the constant ${\cal G}$ depends only on the geometry of the boundary 
$\partial\Omega$ by Eq.\ (\ref{varphi}):
\begin{equation}
{\cal G}=\int_\Omega d{\bf r} \,\nabla\varphi {\hat T}\nabla\varphi
= \int_{\partial\Omega} d{\bf s}\cdot\varphi {\hat T}\nabla\varphi.
\label{G}
\end{equation}

Consider now a two-terminal diffusive wire, so that the surface
$\partial\Omega$ consists of the left $\partial\Omega_L$ and right
$\partial\Omega_R$ contact surfaces, and the open surface $\partial\Omega_0$
with no current through it.  We choose the boundary conditions for $\varphi$
to be
\begin{equation}
\varphi({\bf r})|_{\partial\Omega_L}=0,\; \varphi({\bf r})|_{\partial\Omega_R}=1,
\; d{\bf s} \cdot{\hat T}\nabla\varphi({\bf r})|_{\partial\Omega_0}=0,
\label{boundary}
\end{equation} 
so that $\rho({\bf r})|_{\partial\Omega_L}=\rho_L$, $\rho({\bf
r})|_{\partial\Omega_R}=\rho_R$, $\lambda({\bf r})|_{\partial\Omega_L}=0$,
$\lambda({\bf r})|_{\partial\Omega_R}=\chi$, and $d{\bf s}\cdot{\bf j}({\bf
r})|_{\partial\Omega_0}=0$. Then ${\cal H}$ becomes a function of $\rho_L$,
$\rho_R$, and $\chi$, and the action (\ref{action-stat5}) is the generator of
the cumulants of the transmitted charge. If instead, ${\cal H}$ and $\chi$ are
expressed in terms of ${\cal I}$ (as above for the 1D theory), then one
obtains the logarithm of the distribution of the current,
$\ln[P(I)]=S(I)-tI\chi(I)$, where $I={\cal G}{\cal I}$, according to equations
(\ref{current-density}), (\ref{G}), and (\ref{boundary}). The constant ${\cal
G}$ may be interpreted as a ``geometrical conductance'' of a wire. In
particular, in the ``ohmic'' regime, i.e.\ when $D$ is independent of $\rho$,
we have $I_0={\cal I}(\chi)|_{\chi=0}=D(\rho_L-\rho_R)$, and therefore ${\cal
G}=I_0/[D(\rho_L-\rho_R)]$.  In this case, the ratio $S/I_0$ does not contain
${\cal G}$ and becomes fully universal, proving also the universality of the
result (\ref{ans}) as a special case.

To summarize, we have proven the universality of the FCS of the transmitted
charge for a two-terminal multi-dimensional generalized wire described by the
action (\ref{action-stat4}) with the noise tensor $F(\rho)\hat T$, being
an arbitrary function of the charge density $\rho$, and with the constant
diffusion tensor $D\hat T$. The universality means that the FCS depends
neither on the shape of the conductor, nor on its dimensionality.
\cite{comment} The FCS of a mesoscopic wire given by Eq.\ (\ref{ans}) is a
particular example of  universal FCS. In the more general case, when $\hat
D$ is a function of $\rho$, the FCS depends on the geometry through only one
parameter ${\cal G}$, the geometrical conductance given by Eq.\ (\ref{G}).

\begin{figure}[t]
\centerline{\includegraphics[width=7cm]{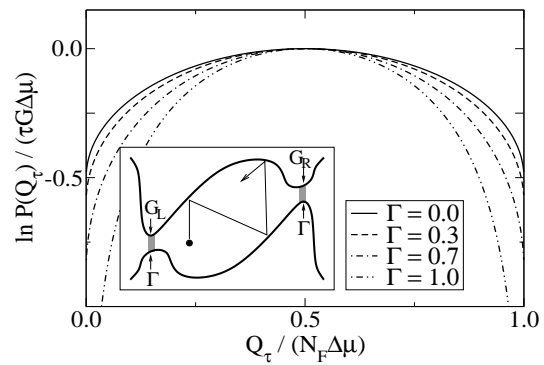}}
\caption{The logarithm of the distribution of charge $Q_{\tau}$ in a symmetric
cavity, $G_L=G_R$, averaged over measurement time $\tau$ in the long time
limit $\tau\gg\tau_D$ and at zero temperature. The results are presented for
several transparencies $\Gamma$ of the point contacts. It is clearly seen that
the tails of the distribution grow in the tunneling limit $\Gamma\ll 1$. The
distribution is symmetric, i.e.\ odd cumulants vanish.  }
\label{Full Statistics}
\end{figure}

\subsection{Charge fluctuations in a chaotic cavity.}
\label{cavityex}

As another example of the applicability of the stochastic path integral
approach, we now consider transport through a chaotic cavity.  This problem
is often investigated in mesoscopic physics because of its simplicity and
conceptual clarity. A cavity consists of a large conducting island of
irregular shape that is connected to two metallic leads through quantum point
contacts (see inset of Fig. \ref{Full Statistics}). 
The distinctive property of the chaotic cavity separating
it from diffusive conductors is that the conductance is determined
solely by the ballistic point contacts.
The chaotic cavity itself may be either disordered or ballistic.
Chaotic cavities can be described by a
semiclassical theory if the point contacts have conductances much larger than
$e^2/h$. The statistics of current flow through the cavity have been addressed
using various methods.  The zero-frequency noise power has been calculated
using random matrix theory\cite{CavityRandomMatrix} and the minimal
correlation principle.\cite{CavityMinCorr} The higher order current cumulants
have been obtained in Refs.\ \onlinecite{us} and \onlinecite{CavityKirill}.
The results are in complete agreement with random matrix theory.

In this section we will address another type of statistics.  In a typical
experimental setup, the cavity is connected to the electrical circuit not only
through the leads, but also through nearby metallic gates via the
electrostatic interaction. Observing potential fluctuations at these
additional gates gives direct insight into the statistics of charge on the
cavity. The noise power of the charge fluctuations in this system has been
calculated in Ref.\ \onlinecite{ChargeNoise}.  The full statistics have been
recently addressed using a random matrix theory.\cite{CavityChargePilgram}
Here, we rederive these results using the stochastic path integral, show
new results on the temperature dependence of these statistics, and also
investigate the instantaneous fluctuation statistics.

In a semiclassical approach, both leads $L,R$ and the cavity are
described by electron distribution functions $f_L,f_R$, and $f$.  The
Fermi functions in the leads $f_{\alpha}=f_F(E-\mu_{\alpha})$ are
characterized by their chemical potential $\mu_{\alpha}$ and their
temperature $T$. The chaotic electron motion inside the cavity makes
the cavity distribution function $f(E,t)$ isotropic and position
independent. Only its energy dependence must be retained.  From now on
we set the electron charge to one, $e=1$. Then the charge $Q$ in the
cavity is given in terms of the electron distribution function and
density of states $N_F$ as $Q= N_F\int dE f$. The average value of
charge is determined by the low-energy cut-off of the integral and is
not relevant for the present discussion. 
The charge and electrostatic
potential of the cavity are related by a geometrical capacitance
$C_g$. In the following, we restrict ourselves to the case $C_g\gg
e^2N_F$ which describes complete screening of the charge in the
cavity.  A more general discussion can be found in Ref.\
\onlinecite{CavityChargePilgram}.  To analyze the time evolution of
the charge, we note that if the size of the cavity is smaller than the
electron-electron and electron-phonon scattering length, every
electron entering the cavity at a certain energy leaves it at the same
energy. The single electron energy is thus conserved and we can
formulate a current conservation law separately for each energy
interval $dE$,
\begin{equation}
N_F \dot{f}(E,t) = J_L(E,t) + J_R(E,t),
\end{equation}
where $J_{\alpha}$ denote ingoing particle currents per energy interval $dE$
in the left and right contacts.  These currents are described by binomial
processes with the cumulant generating function given
by\cite{LevitovStatistics}
\bea
&&
\!\!\!\!\!\!\!
H_{\alpha}(f,i\lambda_{\alpha})dE 
= \Gamma^{-1}G_{\alpha}dE\ln\left[1 + 
\Gamma
f_{\alpha}\left(1-f\right)\left(e^{i\lambda_{\alpha}}-1\right)
\right.
\nonumber\\
&&\qquad\qquad\qquad\qquad+\left.
\Gamma f\left(1-f_{\alpha}\right)\left(e^{-i\lambda_{\alpha}}-1\right)
\right],
\label{H_alpha}
\eea
where we have introduced the conductances of the point contacts $G_{\alpha}$,
$\alpha=L,R$, and their transparency $\Gamma$.

The quantity of interest is the total number of electrons in the cavity
averaged over the measurement time $\tau$,
\begin{equation}
Q_{\tau} = (N_F/\tau) \int\limits_0^{\tau}dt \int dEf(E,t).
\end{equation}
We first consider the long time limit, $\tau\gg\tau_D$, where
$\tau_D=N_F/(G_L+G_R)$ is the average dwell time of an electron in the cavity.
In this limit, the action is stationary with respect to the variables $f$ and
$\lambda$,
\begin{equation}
S = \tau\int dE
\left[H(f,i\lambda)+i(N_F/\tau)\chi f\right],
\quad H=H_L+H_R,
\label{Total Action}
\end{equation}
where the external variable $\chi$ generates the statistics of the desired
quantity $Q_{\tau}$.

At zero temperature $T=0$, the variables $\lambda$ and $f$ are independent of
the energy $E$, and the integration in Eq.\ (\ref{Total Action}) amounts to a
multiplication by $\Delta\mu=\mu_L-\mu_R$. Evaluating the Fourier transform of
the characteristic function $Z(i\chi)$,
\begin{equation}
Z(i\chi) = (2\pi)^{-1}\int dQ\,d\lambda\, \exp(S)\; ,
\label{Z}
\end{equation}
we express the full probability distribution $P(Q_{\tau})$ of charge on the
cavity as an integral
\bea
P(Q_{\tau}) = (2\pi)^{-1}\int d\lambda \exp\left[
\tau\Delta\mu
H(f,i\lambda) \right],\nonumber\\
f=Q_{\tau}/(N_F\Delta\mu). 
\label{Averaged Charge}
\eea
This integral will be calculated in the saddle-point approximation.
For the tunneling limit $\Gamma\ll 1$ and for
open point contacts $\Gamma=1$ we obtain:
\begin{subequations}
\label{result}
\begin{eqnarray}
&&\hspace*{-.9cm}\ln P(Q_{\tau})=\tau G\Delta\mu K(f), 
\\
&&\hspace*{-.9cm}K(f)|_{\Gamma\ll 1}=-\left[\sqrt{f(1-f_0)}-\sqrt{f_0(1-f)}\right]^2,
\label{result1}\\
&&\hspace*{-.9cm}K(f)|_{\Gamma=1}=
f_0\ln\left(\frac{f}{f_0}\right)+
(1-f_0)\ln\left(\frac{1-f}{1-f_0}\right),
\label{result2}
\end{eqnarray} 
\end{subequations}
where $G=G_L+G_R$, and where 
we have introduced the average distribution function
$f_0=G_L/(G_L+G_R)$ in the cavity. We summarize that the results
(\ref{result}) have been obtained under the conditions
$T=0$, $\tau\gg\tau_D$, and for $\Gamma\ll 1$ and $\Gamma= 1$.  These results
can be easily generalized to the case of a multi-terminal cavity.

Although the general case of an arbitrary transparency $\Gamma$ has been also
solved analytically, the final expression for the charge distribution is too
lengthy to be presented here. The Fig.\ \ref{Full Statistics} shows the
distribution $P(Q_{\tau})$ at zero temperature for various transparencies
$\Gamma$ of the point contacts. The cavity is taken to be symmetric $G_L=G_R$.
It is clearly seen that the tails of the distribution grow towards the
tunneling limit.

\begin{figure}[t]
\centerline{\includegraphics[width=7cm]{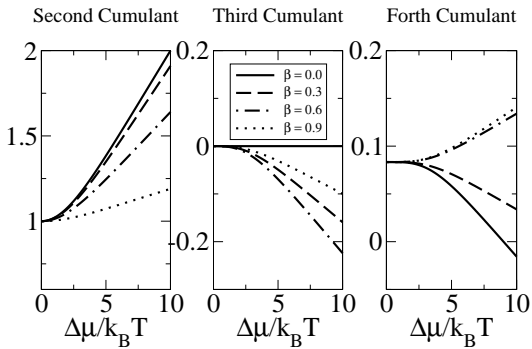}}
\caption{Cumulants of the charge inside a chaotic cavity, $\langle\langle
Q_{\tau}^n\rangle\rangle$, $n=2,3,4$ (in arbitrary units) as functions of the
dimensionless potential difference $\Delta\mu/k_BT$. The parameter
$\beta=(G_L-G_R)/(G_L+G_R)$ characterizes the asymmetry of the cavity.}
\label{Cumulant Plot}
\end{figure}

At finite temperature, further analytical progress can be made by considering
the first few cumulants of the charge $Q_{\tau}$.  The integral (\ref{Total
Action}) for the cumulant generating function has to be evaluated at the
saddle point.  For $\chi=0$ the solution of the saddle-point equations
$\partial S / \partial \lambda = 0$ and $\partial S / \partial f = 0$ are
simply given by $\lambda = 0$ and $f=f_0$, where
$f_0=(G_Lf_L+G_Rf_R)/(G_L+G_R)$ is the average electron distribution function
in the cavity.  From the diagrammatic technique discussed in Sec.\ \ref{cr}
we derive analytical expressions for the first few cumulants. The second cumulant 
has been obtained in Ref.\ \onlinecite{ChargeNoise}.  As an example, we present
here the result for the third cumulant for the case of open point contacts,
$\Gamma=1$:
\begin{subequations}
\begin{eqnarray}
\langle\langle Q_{\tau}^3 \rangle\rangle 
= -\frac{2\tau_D^3}{\tau^2}\,\frac{G_LG_R(G_L-G_R)}{(G_L+G_R)^2}\,
F(\Delta\mu,T),\qquad\\
F(\Delta\mu,T)=
\Delta\mu+ 3\,\frac{\Delta\mu-k_BT\sinh(\Delta\mu/k_BT)}
{\cosh(\Delta\mu/k_BT)-1}\, ,\qquad
\end{eqnarray}
\end{subequations}
where the function $F(\Delta\mu,T)$ is always positive for $\Delta\mu>0$. 
The first few cumulants are plotted in Fig.\ \ref{Cumulant Plot} as a function of
the dimensionless bias $\Delta\mu/k_BT$.  Note that the fourth cumulant may
change its sign as one goes from a symmetric cavity ($\beta=0$) to an
asymmetric cavity ($\beta = 0.9$).

So far we have considered the time of measurement $\tau$ longer than the dwell
time $\tau_D$. Next we consider the opposite limit $\tau\ll\tau_D$ (but still
larger than $\tau_0=\hbar/\Delta\mu$) and study the instantaneous fluctuations
of the charge $Q$ in the cavity at zero temperature, $T=0$. For this purpose
we will use the stochastic path integral (\ref{pathint}) for the propagator
$U(Q_f,Q_i,t)$ of the cavity charge.  The distribution $P(Q)$ of instantaneous
fluctuations can be obtained by taking the $t\to\infty$ limit of the
propagator $U(Q_f,Q_i,t)$ and setting $Q_f=Q$. We note that in the long time
limit, $t\gg\tau_D$, the initial state $Q_i$ relaxes to the stationary state
${\bar Q}$, and as a result the saddle-point expression of the propagator
$U=\exp(S_{sp})$ factorizes according to $S_{sp}=S_0({\bar
Q})+S_i(Q_i)+S_f(Q_f)$.  Here the stationary contribution to the action is
zero, $S_0=0$, since there is no charge accumulation on a long time scale. We
will show that the initial state contribution vanishes, $S_i=0$, so the system
looses its memory about the initial state. Thus we obtain $\ln P(Q)=S_f(Q)$.

We now focus on the case of a cavity with two tunneling contacts ($\Gamma\ll
1$). Using the Hamiltonians in Eq.~(\ref{H_alpha}), and 
replacing the counting variable $\lambda\to i\lambda$, we write the
action as
\begin{subequations}
\label{cav}
\begin{eqnarray}
&&\hspace*{-.9cm}S=G\Delta\mu\int dt [\tau_D\lambda\dot f+h_s(\lambda,f)]\; ,
\label{cav-action}\\
&&\hspace*{-.9cm}h_s=(1-f_0)f(e^{\lambda}-1)+f_0(1-f)(e^{-\lambda}-1)\; ,
\label{cav-Ham}
\end{eqnarray}
\end{subequations}
where $h_s$ is the scaled Hamiltonian.
The saddle point equations take the following form
\begin{subequations}
\bea
\tau_D\dot{f} &=& -(1-f_0)fe^{\lambda}+f_0(1-f)e^{-\lambda},
\label{LF}\\
\tau_D\dot{\lambda} &=& \sinh(\lambda) + (1-2f_0)[\cosh(\lambda)-1].
\label{LC}
\eea
\end{subequations}
The solution of the Eq.\ (\ref{LC}) for $\lambda$ reads
\begin{equation}
\lambda(t) = \ln\left[\frac{1+Af_0\exp(t/\tau_D)}{1-A(1-f_0)\exp(t/\tau_D)}\right],
\end{equation}
where $A$ is the integration constant. 

To show that the initial contribution to the action $S_i$ is zero, we note that
independent of the constant $A$, the absolute value of $\lambda$ is a growing
function with the stationary state given by $\bar\lambda=0$ at $t=-\infty$.
This means that starting from early times $t_0\to-\infty$, the solutions are
$\lambda(t)=0$ and $f(t)-f_0=[f(t_0)-f_0]\exp[-(t-t_0)/\tau_D]$.  They
describe the relaxation of the initial state $f(t_0)$ to the stationary state
$\bar f=f_0$.  Substituting these solutions to Eqs.\ (\ref{cav}) we immediately 
find that $S_i=0$.

After making this point we skip the rest of the details and present the final
result for $\ln P(Q)=S_f(Q)$:
\bea
&&\hspace*{-1.5cm}\ln P(Q)|_{\Gamma\ll 1}=-\tau_DG\Delta\mu\nonumber\\
&&\hspace*{-.1cm}\times
\left[f\ln\left(\frac{f}{f_0}\right) +
(1-f)\ln\left(\frac{1-f}{1-f_0}\right) 
\right],
\label{result3}
\eea
which can now be compared to the results (\ref{result}).
The cumulant generating function for the distribution (\ref{result3}) is given
by $S(\chi)=\tau_DG\Delta\mu\ln[1+f_0(e^{\chi}-1)]$.  Note that
$\tau_DG\Delta\mu=N_F\Delta\mu$ is the total number of the semi-classical
states in the cavity which participate in transport. Therefore the
distribution (\ref{result3}) can be interpreted as being a result of
uncorrelated binomial fluctuations of the Fermi occupations of each
semi-classical state.  We would like to mention that the same result can be
obtained by solving the stationary master equation.

\vspace{1.cm}

\section{Conclusions}
\label{conclusion}

We have put forth a stochastic path integral formulation of
fluctuation statistics in networks.  The mathematical building blocks
of the theory are 1) the probability distributions of transport
processes through the connectors, 2) a continuity equation linking the
connector currents to the charge accumulation in nodes (charge
conservation), and 3) a separation of time scales between nodal
dynamics and connector fluctuations. The relevant action of the path
integral is derived from these considerations and is related to the
probability of (charge conserving) paths in phase space.  The dominant
contribution to the statistics comes from the saddle point
approximation to the path integral, and the generating function for
the interacting system is simply the action at the saddle point.
Fluctuations are suppressed by the number of transporting elementary
charges in the network.  We have considered the continuum limit to
obtain a field theory, and mapped it onto a Langevin equation with
Gaussian noise.  Cascade diagrammatic rules were found in agreement
with Nagaev for the one node case, and extended to general current
correlation functions in an arbitrary network.  Applications to the
current statistics of the diffusive wire and fluctuation statistics of
the charge inside a mesoscopic cavity were also discussed.  As the
building blocks of the theory are classical probability theory, the
potential application of this formalism is very broad and applicable
to any field where fluctuations are important, including mesoscopics,
biology, economics, fluid and chemical dynamics.  \\ {\it Note added
in proof.}---After this paper was submitted for publication, the
authors learned of previous related work by Bertini {\it et
al.}\cite{bertini} Although they did not consider transport
statistics, they did consider the probability to manifest a given
macroscopic fluctuation of the particle density in diffusive lattice
gas models and arrive at the action Eq.~(\ref{act1d}). However, the
Gaussian nature of the local fluctuations was assumed {\it a priori}. We
thank B.\ Derrida for bringing these papers to our attention.

\begin{acknowledgments}
The authors thank M. Kindermann for helpful discussions,
and M. B\"uttiker who collaborated in the initial 
stages of this work.  This work was supported by the Swiss National
Science Foundation, and by INTAS (project 0014, open call 2001).

\end{acknowledgments}


\begin{thebibliography}{02}

\bibitem{BB}
Ya. M. Blanter and M. B\"uttiker, 
Physics Reports {\bf 336}, 1-166 (2000).

\bibitem{b-d}
L. E. Reichl, {\it A Modern Course in Statistical Physics},
(J. Wiley and Sons, New York, 2nd ed., 1998).

\bibitem{examples}
S. M. Bezrukov, A. M. Berezhkovskii, M. A. Pustovoit, and A. Szabo,
J. Chem. Phys. {\bf 113}, 8206 (2000).

\bibitem{gardiner}
C. W. Gardiner, {\it Handbook of Stochastic Methods},
(Springer-Verlag, Berlin, 1990).

\bibitem{LevitovStatistics}
L. S. Levitov and G. B. Lesovik, 
Pis'ma Zh. Eksp. Teor. Fiz. {\bf 58}, 225 (1993)
[JETP Lett. {\bf 58}, 230 (1993)].


\bibitem{Levitov2}
L. S. Levitov, H. Lee, and G. B. Lesovik,
J. Math. Phys. {\bf 37}, 4845 (1996).

\bibitem{us}
S. Pilgram, A. N. Jordan, E. V. Sukhorukov, and M. B\"uttiker,
Phys. Rev. Lett.  {\bf 90}, 206801 (2003).

\bibitem{wire}
H. Lee, L. S. Levitov, and A. Yu. Yakovets,
Phys. Rev. B {\bf 51}, 4079 (1995).

\bibitem{Muzykantskii1} 
B. A. Muzykantskii and D. E. Khmelnitskii,
Phys. Rev. B {\bf 50}, 3982 (1994).

\bibitem{Nazarov1}      
Yu. V. Nazarov, Ann. Phys. (Leipzig) {\bf 8}
Spec. Issue, S1-193 (1999).

\bibitem{Belzig1} W. Belzig and Yu. V. Nazarov, Phys. Rev. Lett. {\bf 87},
067006 (2001).

\bibitem{Bagrets1}
Yu. V. Nazarov and D. A. Bagrets, 
Phys. Rev. Lett. {\bf 88}, 196801 (2002).

\bibitem{Kindermann1} M. Kindermann, Yu. V. Nazarov, and C. W. J. Beenakker,
Phys. Rev. B {\bf 69}, 035336 (2004).

\bibitem{Gefen1}   
D. G. Gutman, Y. Gefen, and A. D. Mirlin, 
in {\it Quantum Noise}, (see Ref. \onlinecite{Nazarov-Blanter});
cond-mat/0210076.

\bibitem{kirillrate}
K. E. Nagaev,
Phys. Lett. A {\bf 169}, 103 (1992).


\bibitem{Beenakker1} 
M. J. M. de Jong and C. W. J. Beenakker in
{\it Mesoscopic Electron Transport}, L. P.
Kouwenhoven, G. Sch\"on, L. L. Sohn eds.,
NATO ASI Series E, Vol. 345 (Kluwer Academic,
Dordrecht 1996).

\bibitem{Langen}
S. A. van Langen and M. B\"uttiker,
Phys. Rev. B {\bf 56}, 1680 (1997).


\bibitem{CavityMinCorr}
Ya. M. Blanter and E. V. Sukhorukov,
Phys. Rev. Lett. {\bf 84}, 1280 (2000).


\bibitem{Nagaev1}       
K. E. Nagaev,
Phys. Rev. B {\bf 66}, 075334 (2002).

\bibitem{CavityKirill}
K. E. Nagaev, P. Samuelsson, and S. Pilgram,
Phys. Rev. B {\bf 66}, 195318 (2002).  

\bibitem{DeJong1}       
M. J. M. de Jong,
Phys. Rev. B {\bf 54}, 8144 (1996).     

\bibitem{derrida}
B. Derrida, B. Dou\c{c}ot, and P.-E. Roche,
cond-mat/0310453.

\bibitem{derrida2}
P.-E. Roche, B. Derrida, and B. Dou\c{c}ot,
cond-mat/0312659.


\bibitem{dp}
M. Doi, J. Phys. A {\bf 9}, 1465; 1479 (1976);
L. Peliti, J. Physique {\bf 46}, 1469 (1985).

\bibitem{cardy}
J. L. Cardy, {\it Scaling and Renormalization in Statistical Physics},
(Cambridge University Press, Cambridge 1996, Ch. 10);
Field Theory and Non-Equilibrium Statistical Mechanics,  
http://www-thphys.physics.ox.ac.uk/users/JohnCardy/

\bibitem{ak}
A. Kamenev, cond-mat/0109316.

\bibitem{multispecies}
Different species of charge may be taken into account by 
letting $Q_\alpha$ be a vector
of these different charge species on each node.  

\bibitem{Nazarov-Blanter} 
{\it Quantum Noise in Mesoscopic Physics},
Proceedings of NATO ARW edited by Yu. V. Nazarov and Ya. M. Blanter,
(Kluwer Academic Publishers, Delft 2003).

\bibitem{kleinert}
H. Kleinert,
{\it Path Integrals in Quantum Mechanics, Statistics, Polymer Physics, 
and Financial Markets}, (World Scientific, Singapore, 3rd ed., 2002).


\bibitem{correlate}
There may be cases where charges in a node are
in fact correlated. This is not difficult 
to incorporate into the formalism, it simply
means that the Hamiltonian in the action
Eq.~(\ref{action1}) will have a more 
complicated dependence on the counting
parameters $\lambda_\alpha$.
For a physical example, see the discussion  
of the hot electron regime in Ref.\ \onlinecite{us}.


\bibitem{oporder}
The ambiguity of at which point to update
the generator is related to the difference between
Ito (beginning) and Stratronovich calculus (midpoint).
\cite{quantization}

\bibitem{Langevin}
J. Zinn-Justin, {\em Quantum field theory and critical phenomena},
Clarendon Press (Oxford, 1993). 

\bibitem{lax}
M. Lax, Rev. Mod. Phys. {\bf 38}, 359 (1966).

\bibitem{freqdep}
K. E. Nagaev, S. Pilgram, and M. B\"uttiker,
Phys. Rev. Lett. {\bf 92}, 176804 (2004).

\bibitem{Oberholzer1}
S. Oberholzer, E. V. Sukhorukov, C. Strunk, 
C. Sch\"onenberger, T. Heinzel, and M. Holland, 
Phys. Rev. Lett. {\bf 86}, 2114 (2001).

\bibitem{Ob2}
S. Oberholzer, E. V. Sukhorukov, C. Strunk, and C. Sch\"onenberger,
Phys. Rev. B {\bf 66}, 233304 (2002).

\bibitem{sbs}
Mathematically, this term is a vector, and because 
the action must be scalar in nature, it must be dotted
into some natural vector of the system.  Physically,
this simply states that at zero bias, there is no preferred
current direction; the retention
of this term demands some broken symmetry in the problem.

\bibitem{cap}
The physical conductance is related to the generalized 
conductance discussed here through a capacitance.

\bibitem{vankampen}
N. G. Van Kampen, {\it Stochastic Processes in Physics and Chemistry},
(North-Holland, Amsterdam 1981).

\bibitem{landau}
L. D. Landau and E. M. Lifshitz, {\it The Classical Theory of Fields},
(Pergamon Press, Oxford, 1962).

\bibitem{white}
Gaussian noise with nontrivial correlations in space and time may be included
by introducing two time and space integrals in the probability
weight Eq.~(\ref{langevin}).

\bibitem{jacobian}
In the process of converting the Langevin equation into
the path integral, there is additionally a functional Jacobian
between $\nabla \nu$ and $\rho$.  This may be written as a determinant
of an operator $M=[\frac{d}{dt}-
\nabla\cdot (\frac{\delta}{\delta \rho}(D \nabla \rho))]\delta(t-t')
\delta({\bf r}-{\bf r'})$.  If $D$ is independent of $\rho$,
this simply alters the overall normalization.
If this is not the case, one may often use the freedom of stochastic 
quantization to render it a constant.\cite{kleinert}
To systemically investigate fluctuations, the determinant
may be written as a fermionic functional integral.\cite{Langevin}

\bibitem{beenakkerexp} 
C. W. J. Beenakker, M. Kindermann, and Yu. V. Nazarov,
Phys. Rev. Lett.  {\bf 90}, 176802 (2003).

\bibitem{reulet}
B. Reulet, J. Senzier, and D. E. Prober,
Phys. Rev. Lett. {\bf 91}, 196601 (2003).


\bibitem{CavitySchomerus}
Ya. M. Blanter, H. Schomerus, and C. W. J. Beenakker,
Physica E {\bf 11}, 1 (2001).

\bibitem{commentrule}
Similarly to the stationary case, the rules may be
formulated with a trivial internal line, but the propagator
$D_{\alpha\beta}(\omega)$ appears in the vertices.

\bibitem{onethird}
C. W. J. Beenakker and M. B\"uttiker, Phys. Rev. B {\bf46},
1889 (1992).


\bibitem{Henny} 
M. Henny, S. Oberholzer, C. Strunk, and C. Sch\"onenberger 
Phys. Rev. B {\bf 59}, 2871 (1999).

\bibitem{Nazarov2}
Yu. V. Nazarov, Phys. Rev. Lett. {\bf 73}, 134 (1994).

\bibitem{Sukhorukov}
E. V. Sukhorukov and D. Loss, Phys. Rev. Lett. {\bf 80}, 4959 (1998); 
E. V. Sukhorukov and D. Loss, Phys. Rev. B {\bf 59}, 13054 (1999). 


\bibitem{commentScr} { If diffusion coefficient $D$ and noise
coefficient $F$ do not explicitely depend on energy (which is usually
the case for metals), it can be shown that the electrostatic potential
along the wire can be absorbed into the electron energy and does not
influence the action\ (\ref{action-stat3}) in the zero frequency
limit.  Finite frequency effects due to fluctuations of the
electrostatic potential are discussed in Ref.\
\onlinecite{ScreeningNagaev}. }


\bibitem{commentT} { It is crucial for the universality of the
statistics that $D(\rho)$ and $F(\rho)$ multiply the same tensor
$\hat{T}$. }

\bibitem{comment}
This proof of the universality can be easily generalized 
to include the coordinate dependence of $\hat D$ and $\hat F$,
in the same way as it has been done for the noise power in 
Ref.\ \onlinecite{Sukhorukov}. 

\bibitem{CavityRandomMatrix}
R. A. Jalabert, J.-L. Pichard, and C. W. J. Beenakker,
Europhys. Lett. {\bf 27}, 255 (1994).

\bibitem{ChargeNoise}
M. H. Pedersen, S. A. van Langen, and M. B\"uttiker,
Phys. Rev. B {\bf 57}, 1838 (1998).

\bibitem{CavityChargePilgram}
S. Pilgram and M. B\"uttiker,
Phys. Rev. B {\bf 67}, 235308 (2003).

\bibitem{bertini}
L. Bertini, A. De Sole, D. Gabrielli, G. Jona-Lasinio, and C. Landim,
J. Stat. Phys. {\bf 107}, 635 (2002); Phys. Rev. Lett. {\bf 87}, 040601 (2001).

\bibitem{quantization}
H. Nakazato, K. Okano, L. Sch\"ulke, and Y. Yamanaka,
Nucl. Phys. B {\bf 346}, 611 (1990). 

\bibitem{ScreeningNagaev}
S. Pilgram, K. E. Nagaev, and M. B\"uttiker,
Phys. Rev. B {\bf 70}, 045304 (2004).



\end{thebibliography}
\end{document}